\documentclass{amsart}
\usepackage{natbib}

\newcommand{\bfs}{{\bf s}}
\newcommand{\bfY}{{\bf Y}}
\newcommand{\bfX}{{\bf X}}
\newcommand{\bfZ}{{\bf Z}}

\title{Bayesian variable selection for spatially dependent generalized linear models}
\author{Kristian Lum\\
Network Dynamics and Simulation Science Laboratory\\
Virginia Bioinformatics Institute\\
Virginia Tech}
\thanks{ I'd like to thank James Johndrow for suggesting and assembling the data set for the election application example, and Melanie Wilson Quintana and Alan Gelfand for helpful discussions. }
\date{} 

\usepackage[utf8]{inputenc} 
\usepackage{textcomp} 
\usepackage{graphicx}  
\usepackage{flafter}  
\usepackage{natbib} 
\usepackage{bbm}
\usepackage{amsmath,a mssymb}  
\usepackage{bm}  
\usepackage{memhfixc}  
\usepackage{pdfsync}  

\newcommand{\bet}{ \mbox{\boldmath $ \eta $} }
\newcommand{\bome}{ \mbox{\boldmath $ \omega $} }

\newcommand{\bbeta}{ \mbox{\boldmath $ \beta $} }

\newcommand{\bzeta}{ \mbox{\boldmath $\zeta$}}

\newcommand{\bepsilon}{ \mbox{\boldmath $\epsilon$}}

\newcommand{\bmu}{ \mbox{\boldmath $\mu$} }

\newcommand{\bgam}{ \mbox{\boldmath $\gamma$} }
\newcommand{\bgamma}{ \mbox{\boldmath $\gamma$} }

\newcommand{\btau}{ \mbox{\boldmath $\tau$} }

\newcommand{\bs}{\textbf{s}}

\newcommand{\bX}{\textbf{X}}

\newcommand{\bp}{\textbf{p}}

\begin{document}
\maketitle
\begin{abstract}

Despite the abundance of methods for variable selection and accommodating spatial structure in regression models, there is little precedent for incorporating spatial dependence in covariate inclusion probabilities for regionally varying regression models. The lone existing approach is limited by difficult computation and the requirement that the spatial dependence be represented on a lattice, making this method inappropriate for areal models with irregular structures that often arise in ecology, epidemiology, and the social sciences. Here we present a novel method for spatial variable selection in areal generalized linear models that can accommodate arbitrary spatial structures and works with a broad subset of GLM likelihoods. The method uses a latent probit model with a spatial dependence structure where the binary response is taken as a covariate inclusion indicator for area-specific GLMs. The covariate inclusion indicators arise via thresholding of latent standard normals on which we place a conditionally autoregressive prior. We propose an efficient MCMC algorithm for computation that is entirely conjugate in any model with a conditionally Gaussian representation of the likelihood, thereby encompassing logistic, probit, multinomial probit and logit, Gaussian, and negative binomial regressions through the use of existing data augmentation methods. We demonstrate superior parameter recovery and prediction in simulation studies as well as in applications to geographic voting patterns and population estimation. Though the method is very broadly applicable, we note in particular that prior to this work, spatial population estimation/capture-recapture models allowing for varying list dependence structures has not been possible. 

\end{abstract}

\section{Introduction}
Variable selection is an especially well-developed topic in statistics, as are methods for incorporating spatial dependence in linear regression and other statistical models. The combination of variable selection with spatial statistics, however, has been left relatively unexplored. In this paper, we unite spatial statistics with variable selection in a broad subset of generalized linear models (GLMs) by allowing each region in an areal model to have its own regression model, i.e. each region's model includes different covariates. We achieve spatial information sharing in variable selection via a prior that induces spatially smoothed inclusion probabilities. Our model for spatially smoothed inclusion probabilities (SSIP) embeds a conditionally autoregressive (CAR) prior on latent standard normal variables arising from the data-augmentation algorithm for probit regression. The probit regression is itself latent, as the binary indicators specify whether a variable is included or excluded from each regional model. By combining this representation of covariate inclusion indicators with any conditionally normal representation of a likelihood, we achieve a fully conjugate Bayesian model with very straight-forward computation. Because many GLM likelihoods can be represented as conditionally normal using data augmentation methods, our approach can be used with many GLMs. We propose a Markov Chain Monte Carlo (MCMC) algorithm for estimation, which also benefits from the full conjugacy of the model, as we are able to marginalize over the regression coefficients to obtain good mixing. Further, we are able to accommodate a richer spatial dependence structure than the one existing method, which requires that spatial dependence structures be representable on a lattice. As such, the existing method is not generally applicable to areal models, whereas our method can be used with any areal model.

The only direct precedent for this work is that of \citet{Smith:2007cr}. They  present a spatial variable selection model for functional magnetic resonance imagine (fMRI) data that also  allows for a different regression model at each location. Their model, however, requires that the spatial locations lay on a grid, a requirement of the Ising model prior they employ (see \citet{Higdon:1994nx} for an overview of the Ising model). Their model inherits the computational difficulties of the Ising model, and they  rely upon approximations to the density and update their  inclusion indicator variables using Metropolis-Hastings. Further,  the parameters of their Ising prior, which control the degree of spatial smoothing, are initially suggested to be pre-specified and not estimated as part of the procedure. They offer a way to simultaneously estimate these parameters which requires gridding the space to approximate to the normalizing constant of the Ising prior.   \citet{Lee:ij} shows an application of this model. \citet{Scheel:2012kl} presents a Bayesian Poisson hurdle model with Ising smoothed variable selection. Similar to \citet{Smith:2007cr}, they use an Ising prior for inclusion probability smoothing. Computation for this model is also cumbersome, as many of the parameters cannot be sampled from known distributions. They rely upon an assortment of methods from the adaptive Metropolis algorithm (\citet{Roberts:2006tg}) to a reversible-jump sampling scheme (\citet{Green:1994hc}). 

Some attention has been paid to spatial variable selection in the genetics literature. \citet{Stingo:2011fv}, for example, model phenotypic outcomes given genotytpes and associated effects on biochemical pathways. They incorporate gene location information in formulating their variable inclusion probability prior. They use a Markov random field (MRF) for the location structure and specify that the probability that a gene is included in the model is dependent upon its neighbors' inclusion. In this case, there is only one regression model, as opposed to our varying site-specific regression models. Neighboring genes on a chromosome then have smoothed probability of being included in the regression model. \citet{Vannucci:2011bs} offers an in-depth review of similar methodologies within the genetics literature.  

\cite{Reich:2010oq} presents a model for variable selection for spatially varying coefficient regression. Here again, there is only one regression model for all (point-referenced) locations, and they select whether each variable should be excluded, included with a  globally constant coefficient, or included with a spatially varying coefficient. 

The use of a latent probit regression to inform prior inclusion probabilities is similar to the model selection approach of \citet{Quintana:qa}. 

The rest of this article is organized as follows. Section \ref{sec:SSIP} introduces the general regression framework for spatially varying models. Section \ref{sec:BinaryModel} introduces a latent variable model for spatially smoothed binary indicators; Section \ref{sec:Gaussian} describes how to incorporate these indicators into a normal linear regression framework to allow spatially smoothed inclusion probabilities with a gaussian likelihood. Subsections of Section \ref{sec:Gaussian} describe the MCMC estimation algorithm, present a simple simulation example, and apply the normal model to election results data. Section \ref{sec:NB} presents a model for negative-binomial regression with the SSIP prior, and discusses the general applicability of the SSIP model to a number of other GLMs. Subsections of Section \ref{sec:NB} propose an estimation algorithm that takes advantage of the conditionally normal representation of the negative binomial distribution using Polya-gamma latent variables, describes how negative binomial regression with the SSIP are a perfect candidate for spatial capture-recapture (CRC), presents a simulation example that shows the increased estimation power gained by leveraging the information contained in all spatial regions simultaneously in CRC, and analyzes a data set of people killed in Casanare, Colombia, aiming to estimate the total number of unaccounted for murders in the region.   Section \ref{sec:conclusion} concludes and offers suggestions for future directions.

\section{Generalized linear model with a SSIP prior} \label{sec:SSIP}

\subsection{Model}
We are interested in developing a method to allow for regionally varying, spatially dependent regression models. For simplicity, we employ a  generalized linear regression model for each region in which,

\begin{equation}
E[\bfY_i]  = g(\bfX_i^T \bbeta_i), 
\end{equation}
 where $\bfY_i$ represents the vector of outcomes in region $i$. $\bfX_i$ is the $m_i \times p$ design matrix associated with region $i$, and $g$ is the link function.  See \citet{1972} for more details about generalized linear models. Borrowing of information for the regression parameters  is permitted through a common prior on each of the $\bbeta_i = [\beta_{i1}... \beta_{ip}]$ vectors of regression coefficients. We employ the spike and slab prior of \citet{MR0359234}, \citet{MR997578}, \cite{Smith:1996fu}, and similar to that used in stochastic search variable selection as in \citet{George:1993zr} (among others) for $\beta_{ij}$ for each $j \in \{1, ..., p\}$.   That is, 
 
 
 \begin{equation} \label{eq:sas}
 \beta_{ij} \sim \gamma_{ij} N(\mu_j, \tau^2_j) + (1-\gamma_{ij})\delta_0(\beta_{ij}),
 \end{equation}
 
  where $\delta_0(\beta_{ij})$ is the Dirac delta function evaluated at $\beta_{ij}$. For any $j$, conditional on $\gamma_{ij}=1$, the $\beta_{ij}$ share a common prior, resulting in sharing of information across regions in estimating the $\beta_{ij}$'s. Put another way, the model has area-specific random effects. Conditional on $\gamma_{ij}=0$, $\beta_{ij}$ has a degenerate distribution at zero, so effectively the $j$th predictor is not included in region $i$'s model. 

The unique feature of the model presented here is that we incorporate spatial structure in the region-specific variable inclusion indicators $\gamma_{ij}$, inducing sharing of information between nearby regions in variable selection. To induce spatial dependence in the $\gamma_{ij}$'s, we specify an independent spatially smoothed inclusion probabilities (SSIP) prior for each covariate inclusion indicator:
\begin{align} \label{eq:SSIP}
&\gamma_{ij} = \mathbbm{1}[ Z_{ij}>0 ] \\
&Z_{ij} \sim N(\rho \frac{1}{n_i}\sum_{k \sim i}Z_k,1),  \nonumber
\end{align}

where $n_i$ are the number of neighbors of region $i$, and the sum is taken across all neighbors of region $i$, as denoted by the $\sim$ relation.By specifying the distribution of $\bgamma_i$ conditional on the latent $\bfZ_k$ of its neighbors, we create spatial dependence in the region-specific models. In other words, if covariate $j$ is included in the regression models for regions that are neighbors of region $i$, then covariate $j$ is more likely to be included in the regression model for region $i$.  The following section includes further discussion of this prior and its properties. 

To complete this model, we place a $N(\mu_0, s_0)$ prior on  the $\mu_j$s and an inverse-Gamma $IG(a_t, b_t)$ prior on the $\tau^2_j$s. 

\subsection{Properties of the SSIP prior} \label{sec:BinaryModel}
Let $\gamma_{ij}$ be a binary variable corresponding to region $i$ which indicates whether the $j$th covariate is included in the model. Following the data augmentation approach of Albert and Chib, we can represent each $\gamma_{ij}$ as a latent Gaussian variable with the binary outcome arising through thresholding, i.e. $\gamma_{ij} = \mathbbm{1}[ Z_{ij}>0 ]$ and  $Z_{ij} \sim N(\mu_{ij}, \tau^2_{j})$. Marginal of $Z_{ij}$, $\gamma_{ij}$ is Bernoulli distributed; the area-specific mean $\mu_{ij}$ controls the parameter of the Bernoulli distribution.  To induce spatial dependence, we place a intrinsically autoregressive (IAR) prior on the latent $Z_{ij}$ (\citet{Besag:1991ys}), $Z_{ij} | Z_{-[i]j} \sim N(\rho \frac{1}{n_i}\sum_{k \sim i}Z_k, \frac{1}{n_i})$ making the model that of equation \ref{eq:SSIP}. 

To understand the implications of this prior on the conditional inclusion probabilities, we notice that marginal of the latent variable $Z_{ij}$,  $Pr(\gamma_{ij}=1 | Z_{-[i]j}) = \Phi(\rho \frac{1}{\sqrt{n_i}}\sum_{k \sim i}Z_k)$, where $\Phi$ represents the standard normal cumulative distribution function. Intuition about the dynamics of this function is perhaps more easily gained if this probability is re-expressed as  $Pr(\gamma_{ij}=1 | Z_{-[i]j}) = Pr(Z_{ij} >0)$ where $Z_{ij} \sim N(\rho \frac{1}{n_i}\sum_{k \sim i}Z_k, \frac{1}{n_i})$.  Holding the average number of neighbors that have the $j$th covariate included constant, as the number of neighbors increases, the probability that the $j$th covariate is included in the $i$th model  increases if more than half of the neighboring models include it  and decreases if more than half of the neighbors exclude it.  If exactly half of the $\gamma_{kj}$'s for region $i$'s neighbors are one, the conditional prior inclusion probability remains one-half. Holding the number of neighbors constant, as the proportion of region $i$'s neighbors that include the $j$th covariate increases, so too does the conditional prior inclusion probability.

If $\rho$ is not restricted to be one (as in the CAR model of \citet{Besag:1974vn}), the joint distribution of $Z_{ij}$ exists and $\bfZ_j \sim N(0, \Sigma)$ where $\Sigma^{-1} = D - \rho W$, $D$ is diagonal, $D_{ii} = n_i$, and $W$ is an adjacency matrix. Under this joint model,  marginally $Z_{ij} \sim N(0,\frac{1}{n_i})$ and the marginal prior probability of inclusion for each covariate is one-half, as the marginal mean is zero. As discussed in \citet{Ley:2009pi}, this has implications for the prior model size; a strategy for relaxing this assumption is discussed in Section \ref{sec:conclusion}.  If $\rho$ is set to be zero, this model collapses to independent spike and slab priors for each region. That is, when $\rho$ is zero, the $\bfZ_j$ are independent and imply that $\gamma_{ij}$ are independent with prior probability, $Pr(\gamma_{ij}=1)=\frac{1}{2}$. 

Under this model, despite having omitted an explicitly spatial distribution on the normal component of the distribution of $\bbeta_j$ of equation \ref{eq:sas}, there is  spatial smoothing of $\bbeta_j$ in expectation:  $E[\beta_{ij}] = (1-Pr(\gamma_{ij}=1))\mu_j$. As $\Pr(\gamma_{ij}=1)$ is a spatially smoothed surface, so too is this expectation.  An explicitly spatial distribution for the normal component of the beta distribution is also possible. For example, one could place an IAR or CAR prior on the subset of the $\bbeta_j$s for which $\bgam_j$ is one-- that is, one could remove the rows and columns of the adjacency matrix for which $\gamma_{ij}=0$. A similar approach to this, without any variable selection, is taken in \cite{Assuncao:2003fk}. The implications of this on the spatial structure could be unexpected, as certain regions would not be included in the spatial smoothing (those where $\gamma_{ij}=0$), and one could produce independent islands in the spatial surface. That is, there might be regions completely separated due to $\gamma$ that are spatially quite close but modeled as spatially independent. More unpredictable still is the spatial structure implied by taking a sub-set of the $\bbeta_j$s where sub-regions are not entirely partitioned off, but rather those $\beta_{ij}$ that are excluded in the spatial model are spaced irregularly. For this reason, we are satisfied with spatial smoothing in expectation and specify an iid prior for the normal mixture component. 

\section{Linear regression with Gaussian error term and the SSIP}\label{sec:Gaussian}
As a first example of the SSIP, we consider a special case of the generalized linear model presented above. Here, we specify a linear regression with Gaussian error term and identity link function. In this case, 
\begin{equation}
\bfY_i = \bfX_i\bbeta_i + \bepsilon_i,
\end{equation}
where $\bepsilon_i \sim N(0,\sigma^2 I)$ has the standard iid error assumption. In this case, the algorithm for estimation is straight-forward.

\subsection{Algorithm}\label{sec:NormAlg}
We use MCMC to fit this nearly fully-conjugate model. In the case where $\rho=1$, the model is fully conjugate and all parameters can be  updated from known distributions. If $\rho$ is treated as a parameter to be estimated, it is the only parameter in the model that lacks a full conditional distribution with known closed form. It can be sampled easily using the Metropolis-Hastings algorithm (see \citet{Chib:1995qf}). The algorithm proceeds as follows.  For each $i$,

\begin{itemize}

\item For each $j$,  sample $[\gamma_{ij}, Z_{ij} |  \bfZ_{-[i]j}, \bgamma_{-[i]j}, \mu_j, \tau_j^2, \sigma_i^2]$ marginally of $\bbeta_i$:
\begin{itemize}
\item First sample $\gamma_{ij}$ marginally of $Z_{ij}$:  $\gamma_{ij} \sim \text{Bernoulli}(p_{ij})$, where 

\begin{equation*} p_{ij} = \frac{ w_{ij} \Psi(\bfY_i | \gamma_{ij}=1)}{w_{ij} \Psi(\bfY | \gamma_{ij}=1) + (1-w_{ij})\Psi(\bfY|\gamma_{ij}=0)}, \end{equation*} 

$w_{ij} = \Phi(\rho \frac{1}{\sqrt{n_i}}\sum_{k \sim i}Z_k)$, and $\Psi(\bfY|\gamma_{ij}=1)$ is the marginal likelihood of the model that includes the $j$th covariate with all the others variables included/excluded according to the current value of $\bgamma_{[-i,j]}$. $\Psi(\bfY|\gamma_{ij}=0)$ is defined analogously. $\Psi$ is the result of analytically integrating out $\bbeta_i$ and is available in closed form. 

\item  Sample $[Z_{ij} |\gamma_{ij}]$:  $Z_{ij} \sim N_{\Omega_{ij}} \left (\rho \frac{1}{n_i}\sum_{k \sim i}Z_{kj}, \frac{1}{n_i}\right)$, where $N_{\Omega_{ij}}$ is a normal distribution truncated to the region $\Omega_{ij}$, $\Omega_{ij} = (-\infty,0]$ if $\gamma_{ij} = 0$, and $\Omega_{ij} = [0,\infty)$ if $\gamma_{ij} = 1$. 

\end{itemize}

\item Jointly update the $i$th region's regression coefficients: $[\bbeta_i | \bgamma_i, \bmu, \btau^2, \sigma_i^2] \sim N(M_i^{-1}m_i, M_i^{-1})$, where $M_i = \frac{\bfX_i^T\bfX}{\sigma_i^2} + \text{Diag}(\btau^{-2})$ and $m_i = \frac{\bfX_i^T \bfY_i}{\sigma_i^2} + \bmu^T \text{Diag}(\btau^{-2})$. 

\item Update $[\frac{1}{\sigma^2_i} | \bgamma_i, \bbeta_i] \sim \text{Gamma}(a + m_i/2, b + (Y_i-X_i\beta_i)^T(Y_i-X_i\beta_i)/2)$. 

\item Update $[\mu_j | \beta_{1j}, ..., \beta_{kj}, \mu, s_0]$ and $[\tau^2_j | \beta_{1j}, ..., \beta_{kj}, a_t, b_t]$ from  normal and inverse-gamma distributions respectively.
\end{itemize}

\subsection{Simulation Example}
We present a simple simulation example to demonstrate the spatial smoothing of regional model covariate inclusion probabilities achieved with SSIP. We simulate a different regression model for each  region on a three by three grid, with up to two covariates plus the intercept included in each model. For each region, we draw a Beta distributed random variable ($\xi_i$) and order them such that the lower numbered regions have the smaller random draws. We then sample the $\gamma$ inclusion indicators with $Pr(\gamma_i=1) = \xi_i$. The result is that lower numbered regions tend to have fewer (or no) covariates included, whereas higher numbered regions do. We repeat this process for each of the two potential covariates to be included in this model.  The $\bfX$ covariate matrix for each region is created by sampling iid uniform random variables, and the regression coefficients that have been determined to be included  in the previous step are sampled as iid $N(5,0.25)$ (first coefficient) and $N(3,1)$ (second coefficient). For each region, we sample four observations from the specified regression model; the variance of the error term is one.  

We run the normal SSIP regression model for 10,000 iterations and compare the results to those obtained by fitting independent regional models using Bayesian Model Averaging (\citet{Hoeting:1999ly} gives a thorough review) as implemented in the R package BMA (\citet{Raftery:2012ve}) and the AIC. In the case of the AIC, inclusion probabilities are not relevant; we select the single model with the lowest AIC. Throughout this article, we will refer to these points of comparison as BMA and AIC. 

\begin{figure}[h]
\begin{center}
\caption{\label{fig:NormEx} Posterior inclusion probabilities for the first covariate (left) and second covariate (right).}
\includegraphics[width=2.5in]{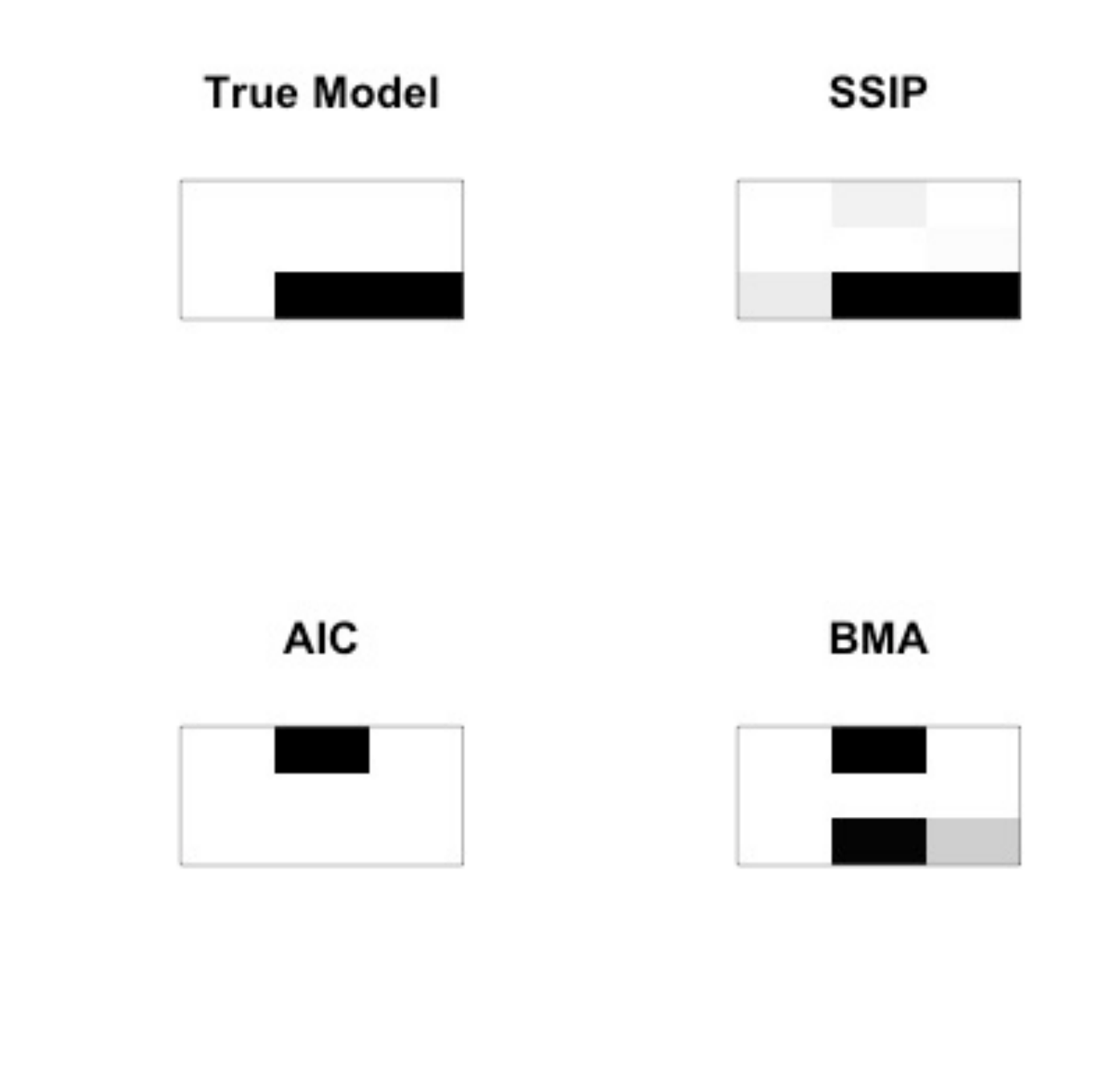}
\includegraphics[width=2.5in]{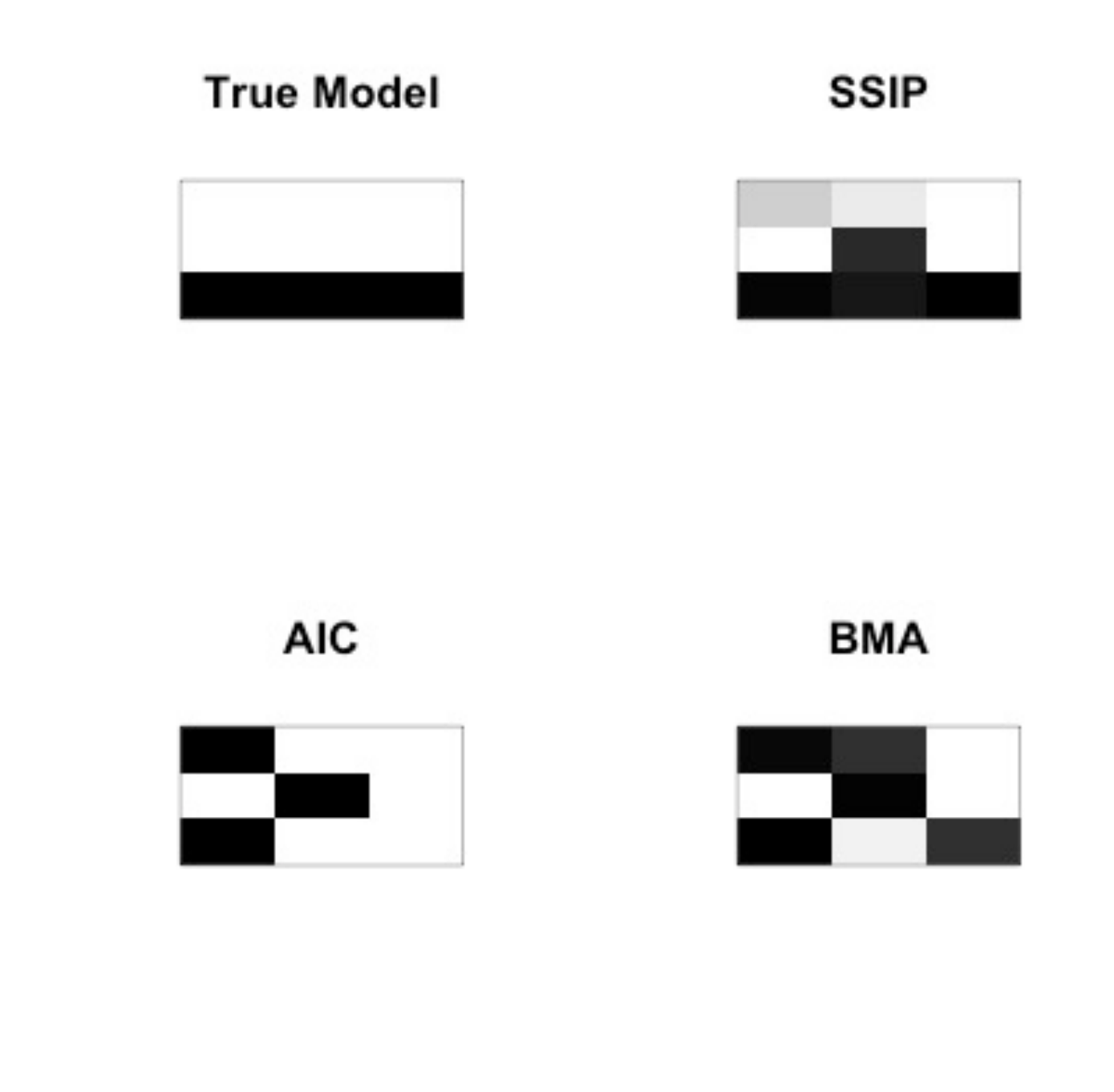}
\end{center}
\end{figure}

Figure \ref{fig:NormEx} shows the inclusion probabilities ranging from one (white) to zero (black) for each region in the simulation for each of the methods in this comparison. The left-most four plots show  the inclusion probabilities for the first covariate. The second covariate is shown on the right. Based only upon a visual inspection, we find that the  SSIP is better able to  identify  the correct model in each region by borrowing information across neighboring regions. BMA and AIC both indicate that the top middle region should exclude the first covariate. However, because all of the neighboring regions include it, the SSIP is able to smooth over them and identify the covariate as likely included in that region's model. 

The second covariate is less precisely determined to be present or absent by all methods due to its lower signal to noise ratio. Again, based upon a visual inspection of these plots, it appears that the SSIP model more accurately captures the pattern of variable inclusion present in this simulation. Table \ref{tab:NormEx} shows the mean squared error between the coefficients'  true value and estimated value. The normal regression model with SSIP produces mean squared errors  significantly lower than the other methods under study. 

These results provide some indication of the advantages of the SSIP method. However the simulation is designed to produce data with high spatial dependence in the area-specific models. It is the only model under consideration which borrows information across all regions. As there are so few observations in each region, it is unsurprising that single-region estimates have high variance and thus difficulty in appropriately identifying regional models. 

\begin{table}[ht]
\begin{center}
\caption{\label{tab:NormEx} Mean squared error of regression coefficients for the SSIP, AIC, and BMA}
\begin{tabular}{rrrr}
  \hline
 & SSIP  & BMA & AIC \\ 
  \hline
MSE & 0.3 & 2.9 & 8.0 \\ 
   \hline
\end{tabular}
\end{center}
\end{table}

\subsection{Election Data}
We consider a data set assembled from the website of the Census Bureau ({http://www.census.gov}). Our dependent variable is the percent of the population that voted Democrat in the 2008 presidential election for every county in the lower 48 states of the United States. As explanatory covariates, we consider the percent of the population that is black, the percent that qualifies as earning income below the poverty line, and the unemployment rate by county. The regression model for election outcomes given demographic covariates is of some practical interest since it provides insight into the factors that influence voting decisions and how these demographic factors differ regionally. For example, the conclusion that poverty and race may be important predictors of voting patterns in some regions and unimportant in others is relevant both for understanding voting results \emph{ex post} as well as for political strategists who wish to tailor campaign materials to specific states or regions. The superior performance of the SSIP method on these data (see below) suggest its usefulness in political science and economics applications. Although in most cases the data would be richer and more complex than in this example, the relative ease with which SSIP scales to larger dimensions presages improved performance on larger datasets.

\begin{figure}[h]
\begin{center}
\caption{\label{fig:PerB} Posterior inclusion probabilities of each variable in the election data example.}
\includegraphics[width=1.5in]{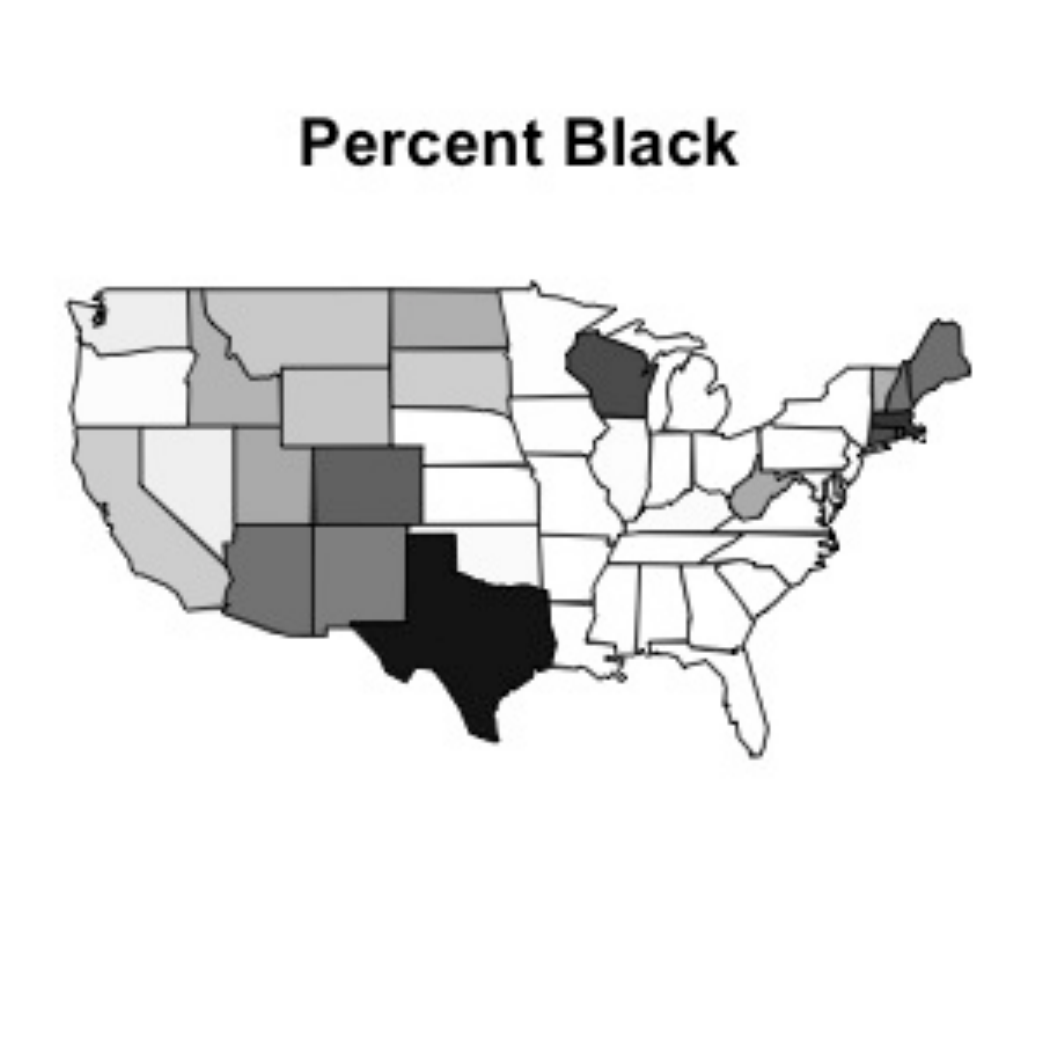}
\includegraphics[width=1.5in]{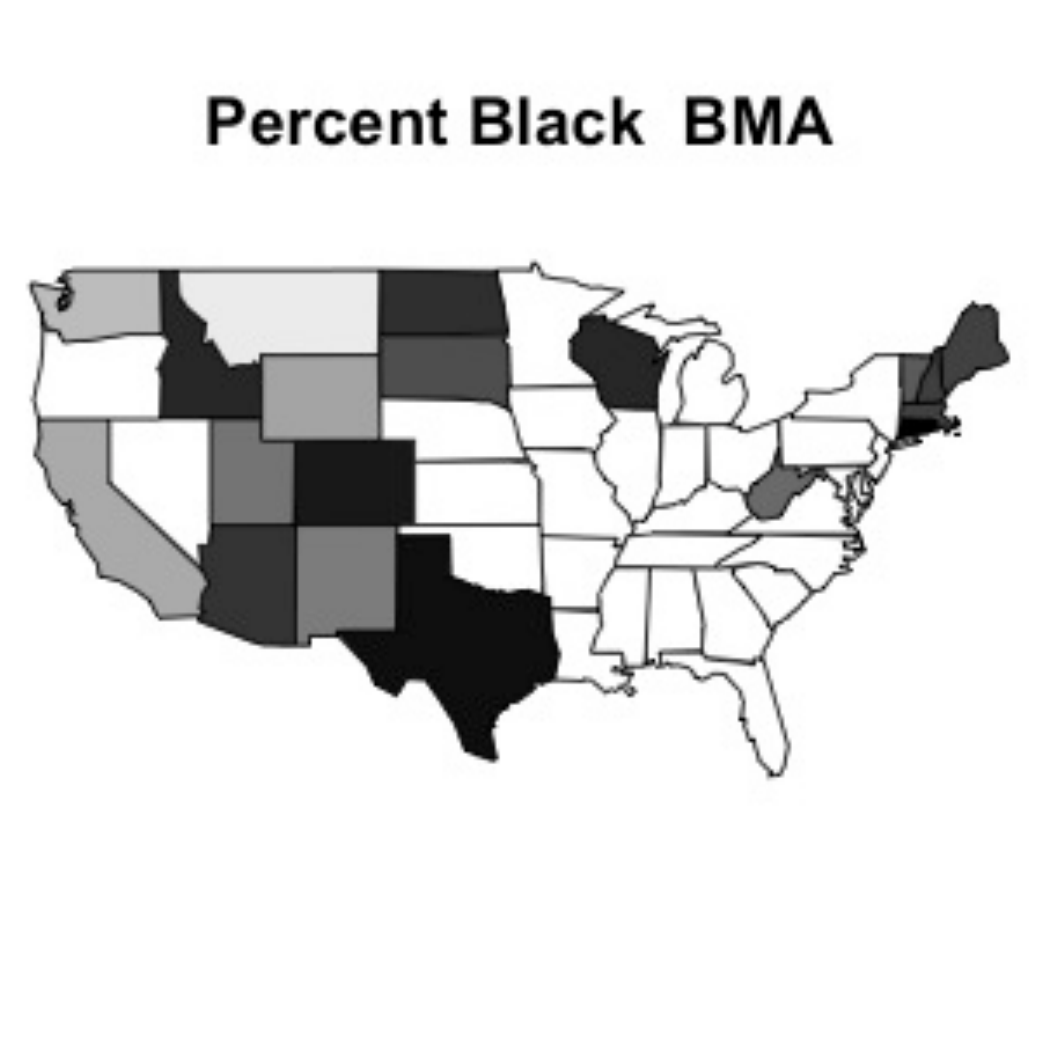}
\includegraphics[width=1.5in]{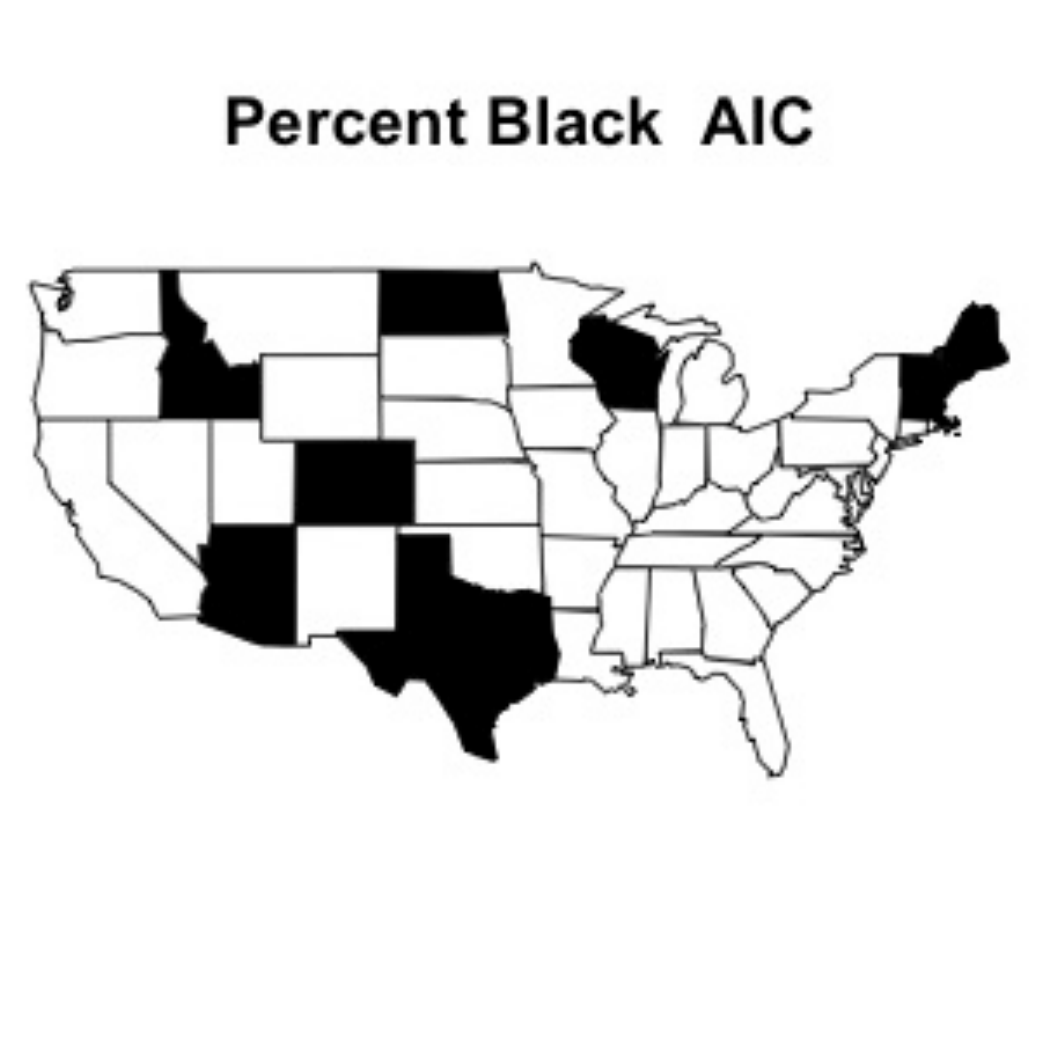}
\includegraphics[width=1.5in]{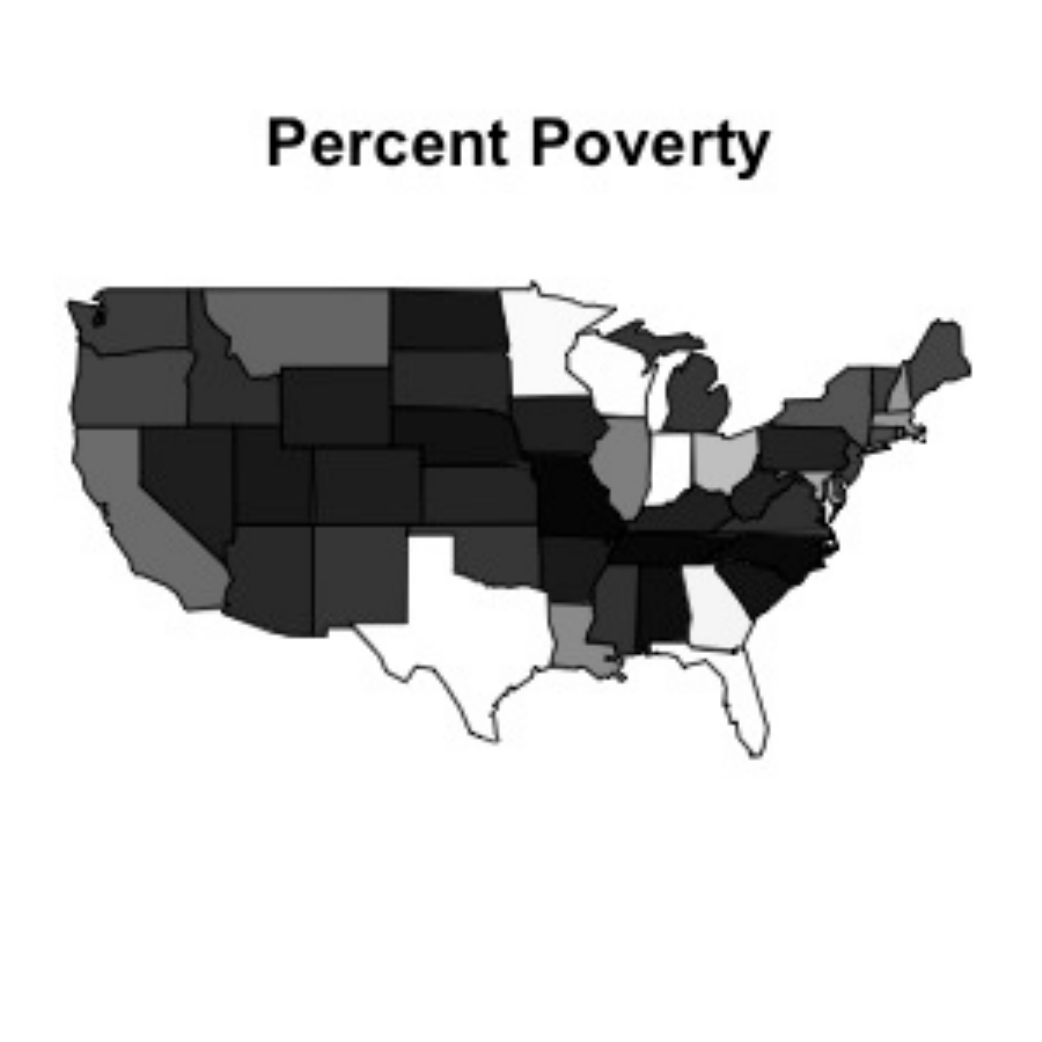}
\includegraphics[width=1.5in]{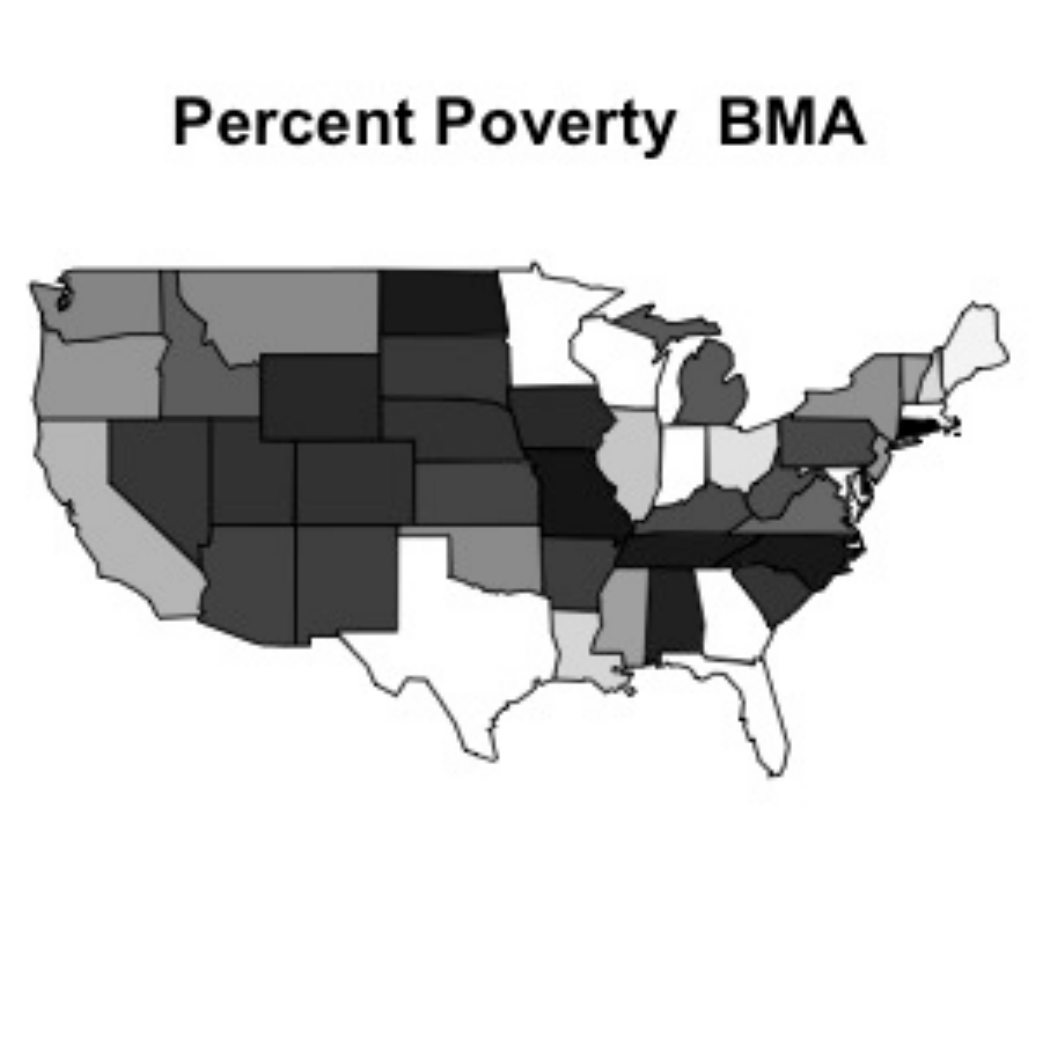}
\includegraphics[width=1.5in]{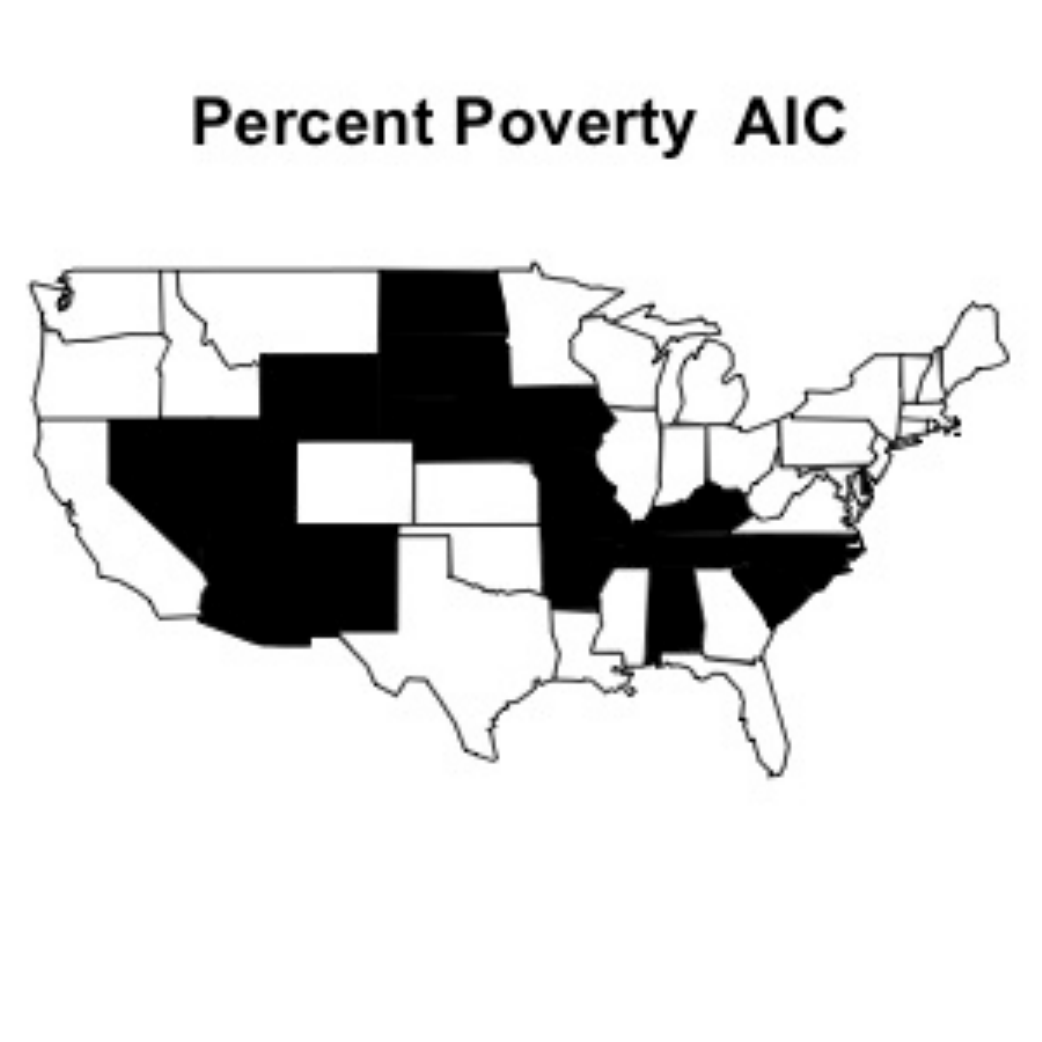}
\includegraphics[width=1.5in]{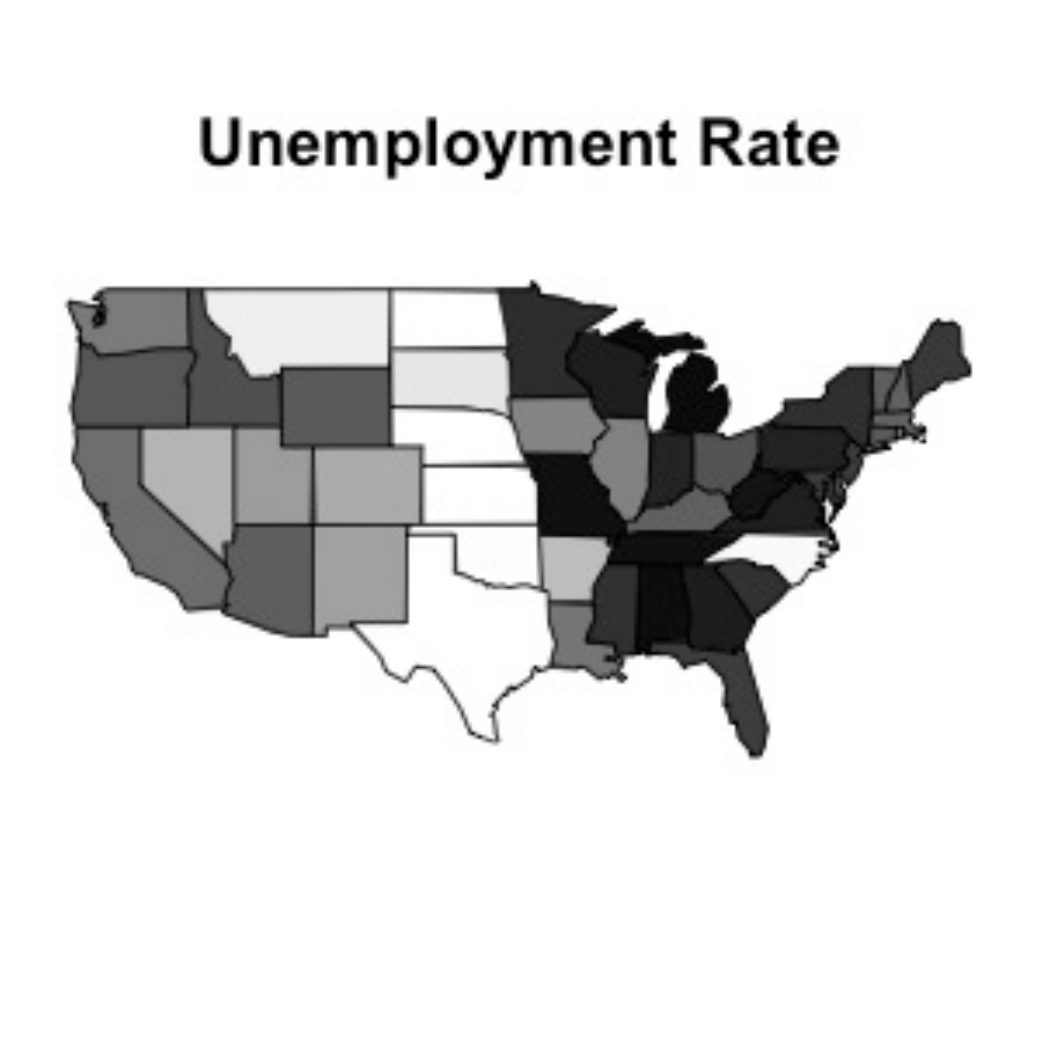}
\includegraphics[width=1.5in]{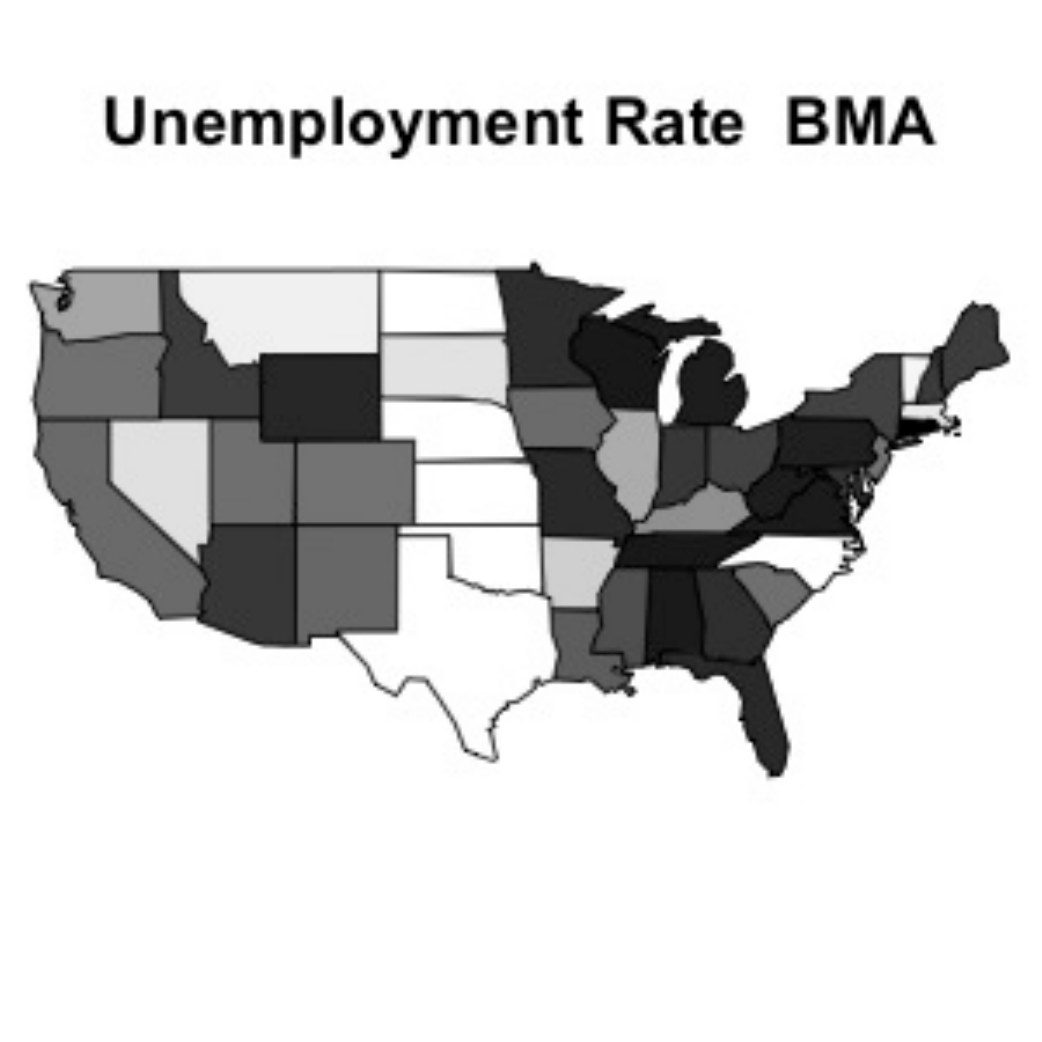}
\includegraphics[width=1.5in]{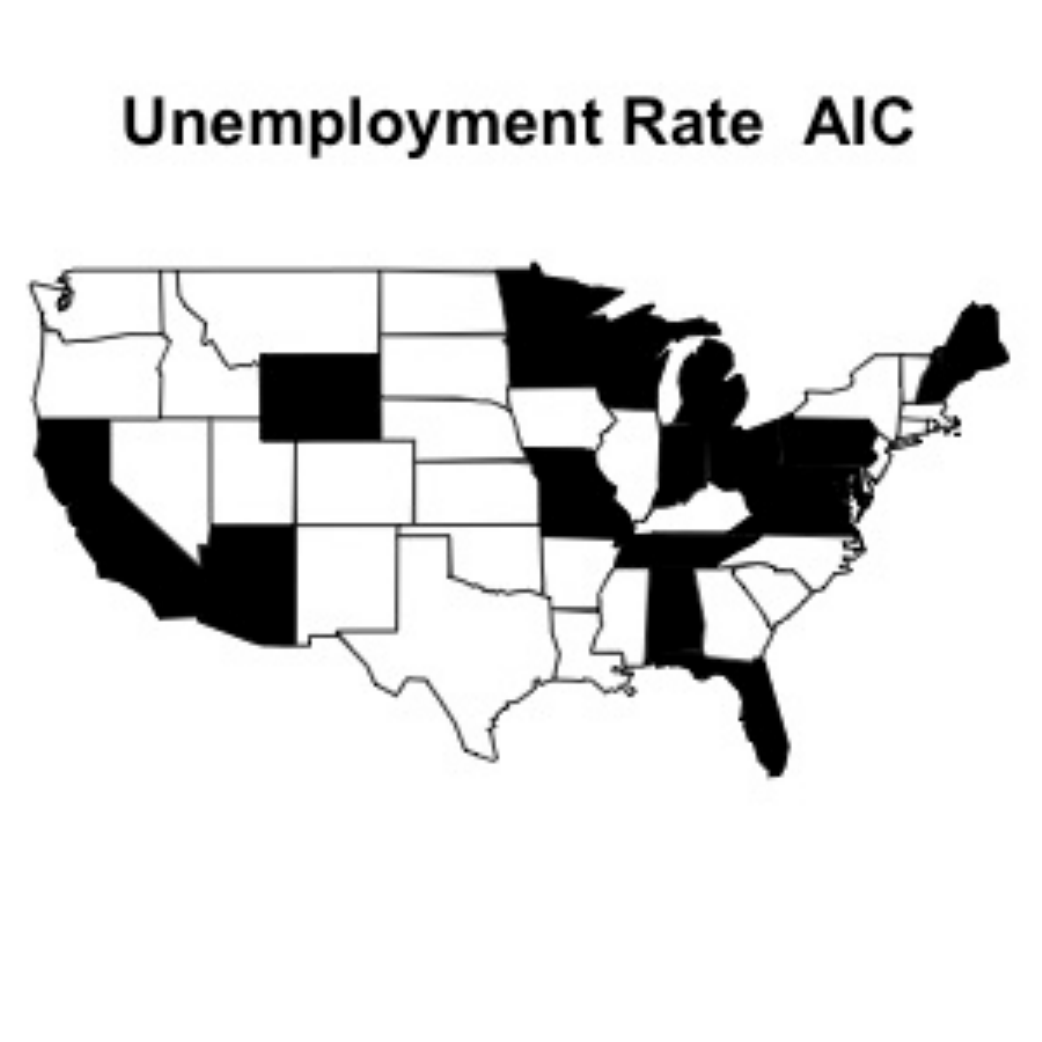}
\end{center}
\end{figure}

We fit the standard linear regression model with spatially smoothed inclusion probabilities as outlined in section \ref{sec:NormAlg}, running the algorithm for 50,000 iterations. As above, we compare the results of our analysis to those obtained by using BMA and the AIC model selection criterion. Results of this analysis are presented in Figure \ref{fig:PerB}.  By using the SSIP, we are able to discern interpretable cultural regions in which different factors are important in explaining the percent of the population that voted Democrat in 2008. For example, focusing on the Percent Black variable, we find that this is a useful predictor (i.e. has high inclusion probability) throughout the South, Mid-West, and Northeast. Both the BMA and AIC methods fail to capture this interpretable regional pattern. Both the SSIP and BMA find that the percent poverty is included in the models of the Gulf Coast and the Great Lakes region. The AIC criterion includes this variable more widely, but a spatial pattern (and corresponding regional interpretation) is noticeably absent. Lastly, the unemployment rate exhibits an interesting spatial pattern of inclusion. East of the Great Plains, there is low probability of inclusion, with the notable exception of North Carolina. The Great Plains region shows a high probability of inclusion, and the region to the west is ambiguous as to whether it should be included or not. Both BMA and the SSIP roughly identify this Great Plains pattern, though the SSIP results in greater spatial smoothing, regional separation, and interpretability. 

\section{Negative binomial regression using Polya-Gamma latent variable representation with SSIP}\label{sec:NB}
Any regression model that admits a conditionally normal representation can easily incorporate the SSIP model for variable selection and maintain conjugate updates. Examples of such models include (Multinomial) Probit regression as in \citet{1993}, Logistic regression as in \citet{Polson:2012uq} and quantile regression as in \citet{Kozumi:2011kx} and \citet{Lum:2012dz}, among others. In fact, any linear model can incorporate the SSIP. If a marginal likelihood is not easily estimated or available in closed form, however, one would have to condition on $\bbeta_i$, which would result in slower mixing. Sampling of $\bbeta_i$ would be subject to the same difficulties inherent in the non-conditionally Gaussian model without the SSIP prior.  Here, we present an example of incorporating the SSIP into negative binomial regression using the Polya-Gamma distribution introduced in the recent work of \citet{Polson:2012uq}. 

We take advantage of the conditionally conjugate representation of the negative binomial distribution presented in \citet{Polson:2012uq} wherein they use the standard  model for over-dispersed count data, 

\begin{eqnarray*}
Y_i  | \lambda_i & \sim & Pois( \lambda_i)\nonumber\\
\lambda_i | h, p_i& \sim & Ga(h, \frac{p_i}{1-p_i}).
\end{eqnarray*}
Marginalizing over $\lambda_i$,  

\begin{equation}
Y_i \sim NB(h, p_i).
\end{equation}

\citet{Polson:2012uq} introduce a latent variable representation of the negative binomial distribution for linear regression, which accommodates fully conjugate updates for both the latent variables and the regression coefficients. They show that the negative binomial likelihood can be represented by an integral, $p(Y_i) \propto e^{\frac{Y_i - h}{2}\psi_i}\int e^{-\omega_i \psi_i^2/2}p(\omega_i)d\omega_i$,  where $p(\omega_i)$ is the Polya-Gamma distribution with parameter $h + Y_i$ and no tilting parameter. From this expression, it is easy to see that conditional upon the latent $\omega_i$, $p(Y_i | \omega_i) \propto e^{-Q(\psi_i)}$. That is, conditionally, the likelihood of $Y_i$ is quadratic in $\psi_i$, which we set equal to $\psi_i = X_i^T\beta$. 

Using this representation, we model areal count data as follows:

\begin{eqnarray*}
\bfY_i | \bbeta_i, \alpha_i & \sim & NB(h, \bp_i)\\
\bp_i & = & \frac{e^{\bX_i^T\bbeta_i}}{1 + e^{\bX_i^T\bbeta_i}}.\\
\end{eqnarray*}
Here, $\bfY_i$ is the vector of counts in region $i$, and $\bp_i$ is the vector of corresponding parameters for the negative binomial distribution. We place the SSIP prior on $\bbeta$ as in Section \ref{sec:SSIP}. 
\subsection{Algorithm}
Following \citet{Polson:2012uq}, we take advantage of the conjugacy of the Polya-Gamma representation of the negative binomial distribution, resulting in another fully-conjugate algorithm. The MCMC algorithm for fitting this model proceeds as follows:

\begin{itemize}
\item Sample all latent $\bome$ random variables from $\bome_i \sim PG(\bfY_i + h, \bX_i^T\bbeta_i)$.
\item Set $\bzeta_i =\frac{\bfY_i - h}{2\bome_i}$. Notice that $\bzeta_i \sim N(\bX_i^T\bbeta_i, \text{diag}(\bome_i^{-1}))$.
\item Proceed as in section \ref{sec:NormAlg}, replacing $\bfY_i$ of section \ref{sec:NormAlg} with $\bzeta_i$. 

\begin{itemize}
\item  For each $j$,
\begin{itemize}
\item Sample $\gamma_{ij} | \bfZ_{-[ij]}\sim \text{Bernoulli}(\bet_{ij})$, with $\eta_{ij} \propto \Psi(\bzeta_i | \bome_i)w_{ij}$.  
\item Sample $Z_{ij} |  \gamma_{ij}, \bfZ_{-[ij]} \sim N_{\gamma_{ij}}(\rho \sum_{k\sim i}Z_{kj}, n_i^{-1})$. 
\end{itemize}

\item Jointly update $\bbeta_i | \bgamma_i, \bmu, \btau^2$. 
\item Update $\bmu$ and $\btau^2$. 
\end{itemize}
\end{itemize}

\subsection{Capture-recapture}
Capture-recapture (CRC) is a natural application of negative binomial regression with SSIP. CRC seeks to estimate the total size of a population based upon multiple sub-samples of the population and the overlaps among them. Let there be $k = \{1, ..., K\}$ sub-samples (or lists) collected. Then, the data consists of counts of the number of individuals that appear on each list intersection. We define a list intersection as a binary string, $d_1... d_K$, where $d_k=1$ for an intersection including list $k$ and 0 otherwise.  $Y_{d_1... d_K}$ is the number of individuals who appear in intersection $d_1... d_K$. This is perhaps most easily explained by example.  $Y_{001}$ is the number of elements that appear only on list three when $K=3$;  $Y_{1010}$ is the number of elements that appear on lists one and three but do not appear on lists two and four for $K=4$. The traditional method  (see \citet{Fienberg:1972bh}) then  fits a Poisson regression to the above counts, 

\begin{eqnarray}
Y_{d_1...d_K} & \sim & \text{Pois}(\mu_{d_1...d_K}) \nonumber\\
\log(\mu_{d_1...d_K}) & = & \alpha + \beta_{1}d_1 + ... \beta_{K}d_K.\label{eq:loglin}
\end{eqnarray}

The number of elements seen on no list ($Y_{0...0}$) is then estimated as $e^\alpha$. As specified above, this model assumes that the lists are collected independently.  Incorporating list dependence is as simple as including interaction terms to the log-linear regression, as first pointed out by \citet{Fienberg:1972bh}. For example,  to account for dependence between list one and two, one would also include $\beta_{12} d_1 d_2$ in Equation \ref{eq:loglin}. 

CRC generally relies upon the assumption that all members of the population have equal probability of being sampled by each list. When it is unreasonable to believe that this is true, stratification is often done to break  the samples into sub-populations within which it is plausible that the probability of capture for each member of the sub-population is approximately equal by list (\citet{Bruno:1994dq}). One variable on which to stratify is spatial location. Particularly in human applications of population estimation, in which people in different geographic or administrative regions may have differing probability of being reported to each data collecting agency, it is sensible to subdivide the population spatially. Additionally, it is also often of interest to have separate estimates by region, making spatial stratification useful both as a tool to satisfy the model's assumptions and as a way to achieve a higher resolution analysis of the spatial distribution of unreported individuals.  However, the degree to which stratification is possible is dependent upon the amount of data available. If one stratifies too much,  each stratum is left with so little data pertaining to it that independently obtaining reasonable estimates for individual strata is impossible. 


Estimation of these models is further complicated by the fact the there is often little information {\it a priori} about which model ought to be used, i.e. which interaction indicators between lists should be included. However, it is logical that spatially proximate regions should have similar models, as collection agencies may operate similarly regionally and sentiment that would result in a population trusting an agency with its data may also exhibit a spatial patterns.  This makes capture-recapture a perfect candidate for using SSIP, which will allow for finer spatial stratification through borrowing of information across space, both in terms of the mean estimates and informing about variable inclusion probabilities. 

Recent work of \citet{Royle:2008uq} and \citet{Marques:2012kx} also presents a spatial capture-recapture model. \citet{Royle:2008uq} considers a continuous spatial resolution (often not available in human applications due to privacy concerns)  in which they assume no list dependence. In our case, list dependence is absolutely required, and in fact, varying spatial models are necessary as well. The spatial aspect of this work is due to the fact that animals may move during the course of the study.  As we are considering a retrospective study of violence, victims can only be killed at one place in time and space thus making these models inappropriate in this context.

\subsection{Simulation Example}
We  consider a case in which there are four lists (K=4) which sample individuals that exist at spatial locations contained in  $\mathcal{S} = [0,1]^2$. Let the intensity surface for the point distribution, $\lambda(\bs)$, be given by $\lambda(\bs) = c \left (\frac{\sqrt{2}}{2} -|| \bs - .5|| \right )$. Figure \ref{fig:lam} shows  this intensity surface and one realization of the locations of individuals sampled from this process. Our first simulated list can detect an event at location $\bfs = \{x, y\}$ with probability $Pr(L_1(x,y)= 1) = \frac{1}{8}(x + y)^3$. That is, for each location generated by $\lambda(\bs)$, it is included in $L_1$ with spatially varying probability described by a cubic in $x$ and $y$. In order to induce list dependence in some spatially adjacent regions, list two samples an event with probability $Pr(L_2(x,y) = 1 | L_1(x,y)) = \frac{3}{16}y^3 + \frac{1}{16} + \frac{1}{2}\mathbbm{1}[L_1(x,y)=1] \mathbbm{1}[x < 0.4]$. We make a similar conditional statement for $L_3$; $Pr(L_3(x,y)=1 | L_1(x,y)) = \frac{3}{16}x^2 + \frac{1}{16} + \frac{1}{2}\mathbbm{1}[L_1(x,y)=1] \mathbbm{1}[x > 0.4]$. Ergo, on the left-hand side of the space, those events that were detected by list one have a higher probability of being detected by list two, whereas on the right-hand side of the space, there is positive list dependence between list one and list three.  List four is generated independently of all other lists, $Pr(L_4(x,y)=1) = \frac{3}{16}x^2 + \frac{1}{16}$. In summary, we sample locations according to a non-homogeneous Poisson process in continuous space. At each of these locations, each of the four lists detects individuals with the spatially varying probabilities and spatially varying dependence. This structure does not precisely simulate from our negative binomial regression model, though it does induce the properties that exist in real applications and which our model addresses.  We discretize the space into a $5 \times 5$ grid of cells to create the areal units.


We fit the negative binomial regression model with SSIP as shown above with one small difference: we place a CAR prior on the regional intercepts. That is, we specify that $\bp_i  =  \frac{\exp[\alpha_i + \bX_i^T\bbeta_i]}{1 + \exp[\alpha_i + \bX_i^T\bbeta_i]}$, where $\bX_i$ are the intersection indicators as above and $\alpha_i \sim CAR(\tau_{\alpha})$. Even with this added complexity, we retain fully conjugate updates for all parameters.

We take a Bayesian model averaging approach to estimating the  total number of un-sampled individuals. That is, for region $i$, we use the ``model average" estimate (\citet{Raftery:1997kl}), $\hat{\alpha_i} = \frac{1}{M} \sum_{m=1}^M \alpha_i^{(m)}$,  i.e. the posterior mean estimate over all iterations of $\alpha_i$ averaged over all of the sampled models. By this notation, we mean that we run the algorithm for $M$ iterations,$m$ indexes iteration, and $\alpha_i^{(m)}$ is the sampled value at iteration $m$ of $\alpha_i$.

To investigate what is gained by including both spatial smoothing of the response surface and spatial smoothing of the inclusion probabilities, we select implementations of models for comparison that include some of these features. Our first two points of comparison are independently fit BMA and AIC estimates as before. An additional point of comparison is a fitting a spatial model with R-INLA (\citet{Rue:2009uq}), an implementation of nested Laplace approximation (\citet{Rue:2009mi}). Using R-INLA,  we  estimate a model with a spatial IAR intercept and common \emph{iid} normal prior for each of the other regression coefficients. Essentially, we use the same model as we will fit with the SSIP but the model does not vary regionally. We include all covariates  and let the random effects determine values for each of them. Ideally, this would be able to estimate non-existent effects as zero coefficients. 

Table \ref{tab:simtab} shows the results of this experiment. We find that if we do not share information among all of the strata, there are many regions for which no reasonable estimate can be produced. This is demonstrated by the fact that in the methods that do not leverage the shared information, there are several estimates with infinite confidence intervals. In these intervals,  some of estimates were on the order of $10^{13}$.  By comparing our method to the R-INLA implementation, we see the added value of allowing  different models by region. Our method has smaller credible intervals than those produced by the other models. In addition, our model has empirical coverage rate approximately equal to the nominal coverage rate of $0.95$,  it has the lowest root mean squared error as applied to the difference between the true number of unsampled individuals and the estimated number, and the mean median error is also the lowest among the methods considered. 

\begin{figure}[h]
\begin{center}
\caption{\label{fig:plot1}Comparison of fitted values for synthetic MSE example.}
\includegraphics[width=5in]{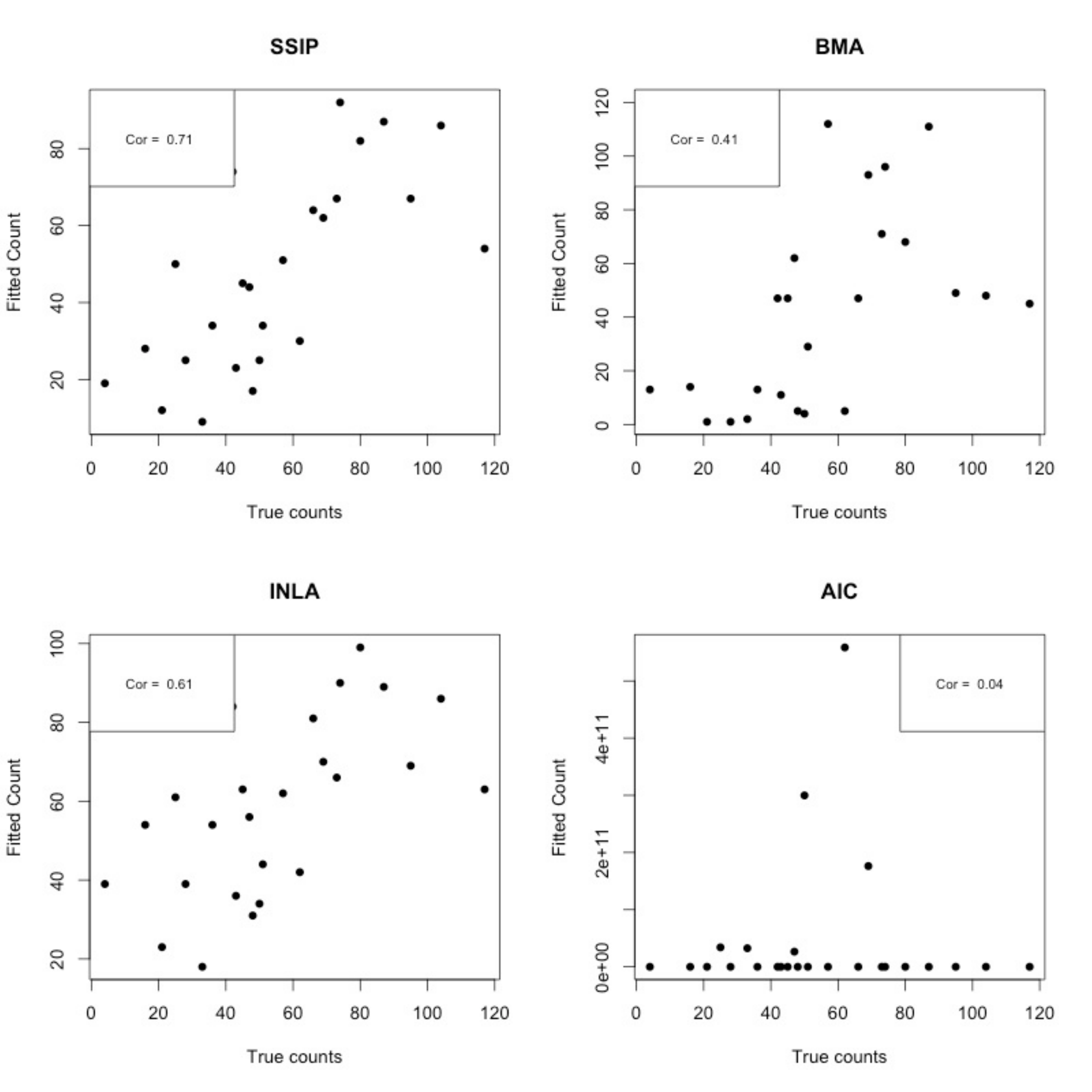}
\end{center}
\end{figure}

\begin{figure}[h]
\begin{center}
\caption{\label{fig:gamma12}Comparison of inclusion probabilities for list interaction $\beta_{12}$ by region.}
\includegraphics[width=3.5in]{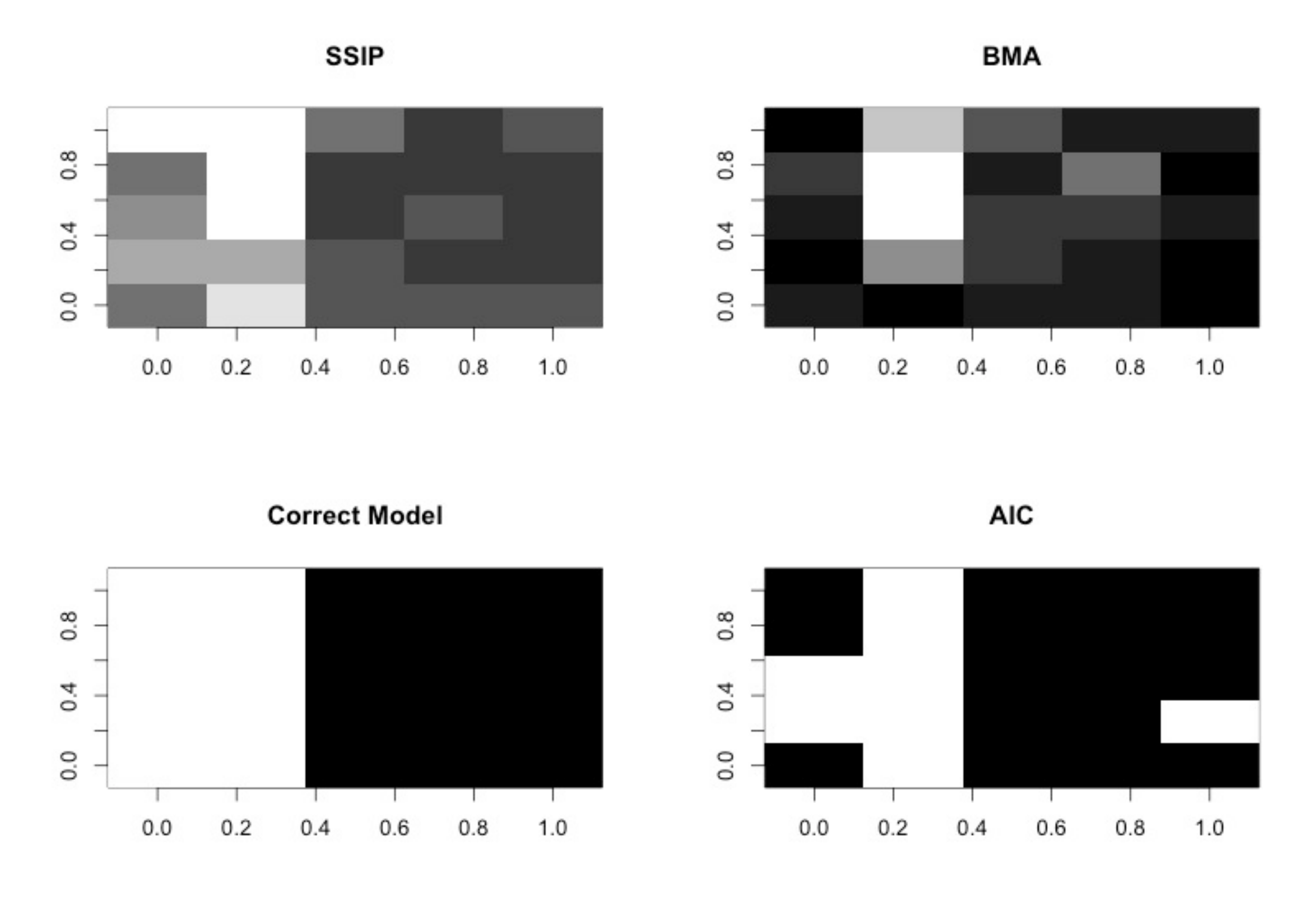}
\end{center}
\end{figure}

\begin{figure}[h]
\begin{center}
\caption{\label{fig:gamma13}Comparison of inclusion probabilities for list interaction $\beta_{13}$ by region.}
\includegraphics[width=3.5in]{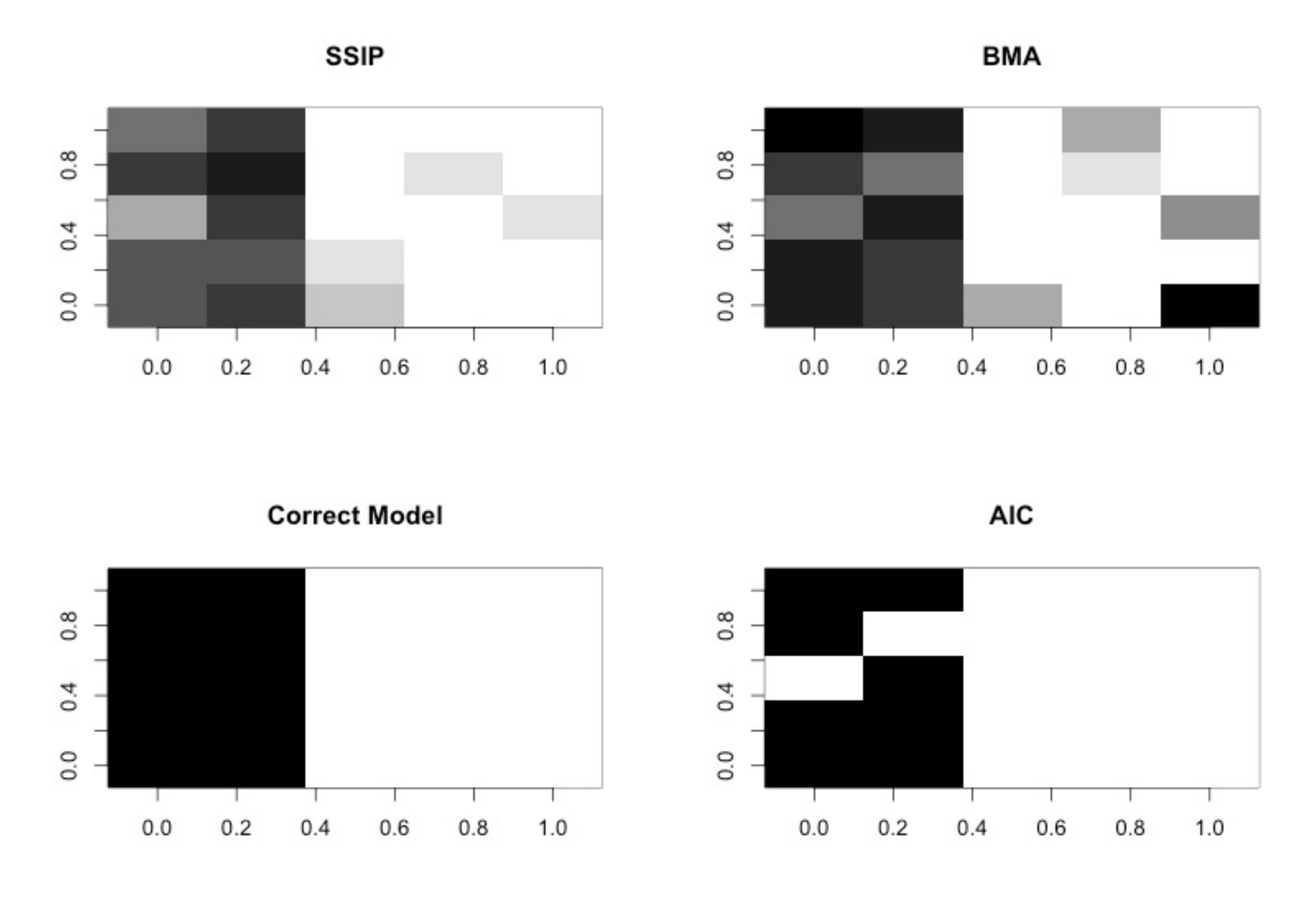}
\end{center}
\end{figure}

Figure \ref{fig:plot1}  illustrates this point by showing a comparison of the true counts to the estimated counts produced by each method. Our method produces a correlation between the true and estimated values of 0.71, whereas the next best method achieves only 0.61. Figure \ref{fig:gamma12} shows the inclusion probabilities for the interaction between list one and list two. Figure \ref{fig:gamma13} shows the same for the interaction between list one and list three.  It should be noted that the AIC criterion performs admirably in terms of variable selection. However, it fails to produce reasonable population estimates (and also produces infinite confidence intervals) in many regions. BMA appears to have less success at identifying the correct regional models, though all of the estimates in this particular simulation are of a reasonable order of magnitude and have finite credible intervals. The SSIP is more successful than BMA at identifying the regional model pattern and less successful than the AIC. Again, this drawback of the SSIP is outweighed by its superior performance in terms of predicting the number of uncounted individuals and smaller and more accurate credible intervals.

\begin{table}[ht]
\begin{center}
\caption{\label{tab:simtab}Results of MSE simulation example. }
\begin{tabular}{ccccc}
  \hline
  True Count & SSIP & BMA & INLA & AIC \\ 
  \hline
 33 & [ 1 , 32 ] & [ 0 , 12 ] & [ 4 , 48 ] & [ 0 , Inf ] \\ 
 50 & [ 4 , 75 ] & [ 1 , 10 ] & [ 10 , 81 ] & [ 0 , Inf ] \\ 
   62 & [ 5 , 93 ] & [ 1 , 21 ] & [ 13 , 100 ] & [ 0 , Inf ] \\ 
  51 & [ 8 , 95 ] & [ 0 , Inf ] & [ 14 , 101 ] & [ 6 , 66 ] \\ 
   21 & [ 1 , 40 ] & [ 1 , 1 ] & [ 6 , 60 ] & [ 0 , Inf ] \\ 
  48 & [ 3 , 51 ] & [ 0 , 29 ] & [ 9 , 73 ] & [ 3 , 107 ] \\ 
   95 & [ 16 , 189 ] & [ 3 , 772 ] & [ 25 , 152 ] & [ 8 , 228 ] \\ 
   117 & [ 16 , 135 ] & [ 11 , 183 ] & [ 24 , 137 ] & [ 14 , 108 ] \\ 
  73 & [ 20 , 165 ] & [ 11 , 457 ] & [ 25 , 142 ] & [ 21 , 574 ] \\ 
   43 & [ 5 , 67 ] & [ 1 , 85 ] & [ 11 , 84 ] & [ 4 , 109 ] \\ 
   69 & [ 14 , 176 ] & [ 7 , 1094 ] & [ 24 , 159 ] & [ 0 , Inf ] \\ 
   87 & [ 23 , 228 ] & [ 16 , 745 ] & [ 33 , 191 ] & [ 39 , 508 ] \\ 
   104 & [ 25 , 203 ] & [ 9 , 234 ] & [ 34 , 179 ] & [ 23 , 141 ] \\ 
   80 & [ 30 , 182 ] & [ 30 , 154 ] & [ 41 , 202 ] & [ 33 , 128 ] \\ 
   45 & [ 14 , 117 ] & [ 9 , 238 ] & [ 23 , 137 ] & [ 16 , 77 ] \\ 
   57 & [ 12 , 137 ] & [ 16 , 773 ] & [ 21 , 140 ] & [ 28 , 630 ] \\ 
   74 & [ 28 , 224 ] & [ 28 , 326 ] & [ 35 , 192 ] & [ 41 , 259 ] \\ 
   66 & [ 21 , 154 ] & [ 13 , 158 ] & [ 33 , 168 ] & [ 14 , 82 ] \\ 
   42 & [ 24 , 172 ] & [ 14 , 154 ] & [ 34 , 172 ] & [ 29 , 125 ] \\ 
  16 & [ 7 , 78 ] & [ 2 , 71 ] & [ 19 , 119 ] & [ 8 , 121 ] \\ 
   28 & [ 4 , 76 ] & [ 1 , 1 ] & [ 11 , 97 ] & [ 7 , 566 ] \\ 
   47 & [ 10 , 125 ] & [ 7 , 502 ] & [ 20 , 126 ] & [ 0 , Inf ] \\ 
   36 & [ 10 , 81 ] & [ 4 , 38 ] & [ 20 , 114 ] & [ 10 , 46 ] \\ 
   25 & [ 13 , 132 ] & [ 18 , 772 ] & [ 21 , 137 ] & [ 0 , Inf ] \\ 
  4 & [ 4 , 57 ] & [ 1 , 98 ] & [ 11 , 93 ] & [ 5 , 1507 ] \\ 
   Empirical Coverage & 0.96 & 0.72 & 0.92 & 0.64 \\ 
   RMSE & 107 & 190 & 113 & 660177792443 \\ 
   RMSE - no outliers & 107 & 190 & 113 & 190 \\ 
   Mean Median Difference & 16 & 30 & 18 & 45103972194 \\ 
   \hline
\end{tabular}
\end{center}
\end{table}

\subsection{Application: Estimating the number of killings in Casanare, Colombia}
 CRC is particularly appropriate for an application in the field of human rights because the total number of abuses is often under-reported-- whether it is due to deliberate attempts by the perpetrators to keep their crimes hidden, fear of retribution by family members who would normally report such crimes, lack of an infrastructure for reliable documentation, or any of numerous other reasons. In studies of violence, it is often not the case that the lists are independent. For example, one group documenting violations may have a policy of referring victims or family members of victims to another agency. Despite the fact that we know it is probable that there are list interactions, it is very difficult to specify a model with the appropriate interaction terms included {\it a priori}. As data collecting agencies may operate differently regionally and are trusted with varying degree, it is appropriate to allow the model to vary by location. 
 
  We analyze a data set of the number of individuals killed in Casanare, Colombia between 2001-2004\footnote{This data is made  freely available by Benetech's Human Rights Data Analysis Group (HRDAG) at {https://www.hrdag.org/about/CasanareSummaries.html}},  a time period during which the region was in the midst of a bloody conflict involving battles between paramilitaries, guerrillas, and the Colombian military (\citet{Guberek:2010fk}). In prior analyses of this data, it was not possible to estimate the number of undocumented killings in each municipality and each year separately because of the sparsity of the data collected. In one analysis, \citet{:cr} aggregated all of the southern municipalities (Sabanalarga, Villanueva, Monterrey, Aguazul, Tauramena, Man\'i, Chameza, Recetor and Yopal)  to get an estimate of the combined total number of killings in the south for two nested time periods, 2001-2004 and 1998-2005. The later analyses of \citet{Guberek:2010fk} and \citet{Lum:2010fk} used slightly less regional aggregation, grouping the area into four regions: Center (Yopal and Augazul); Piedemonte (Sacama, La Salina, Tamara, Recetor, Chameza, and Nunchia); South(Tauramena, Monterrey, Villanueva, Man\'i, and Sabanalarga); and the Plains (Hato Corozal, Paz de Ariporo, Por\'e, San Luis de Palenque, Trinidad, and Orocue). They were, however, able to temporally disaggregate the analysis to produce yearly estimates. The necessity of the spatial aggregation was due to the inability of the available models to reliably produce estimates in a sparse data setting. 
 
 In order to borrow information across years, we include a latent AR(1) shift in the intercept. In this case, $\bp_{it}  =  \frac{ \exp[\alpha_i + \zeta_t + \bX_{i}^T\bbeta_i]}{1 + \exp[\alpha_i + \zeta_t + \bX_{i}^T\bbeta_i]}$, where $t$, which indexes time,  ranges from 2001 to 2004. For the temporal effect, $\zeta_t$, we place an AR(1) prior. Again, despite this added complexity, because of the conditional normality of the components in this model, both $\alpha_i$ and $\zeta_t$ can easily be updated from known full conditional distributions (multivariate normal). For details, see \citet{Prado:2010ff}.  Notice that the region-specific model and parameters remain constant  for all years within a region.  Using our modified model, we obtain estimates of the number of unreported murders from 2000-2004 for each individual municipality and year. 
 
 We include data from  five administrative lists  and include up to four-way interaction terms as potential covariates, resulting in 31 potential covariates for each region. We enforce the restriction that main effects and an intercept must be present, leaving 20 covariates for the algorithm to include or exclude appropriately. The results of the SSIP are shown in Table \ref{tab:coltab}. 
 
\begin{figure}[h]
 \includegraphics[width = 2in]{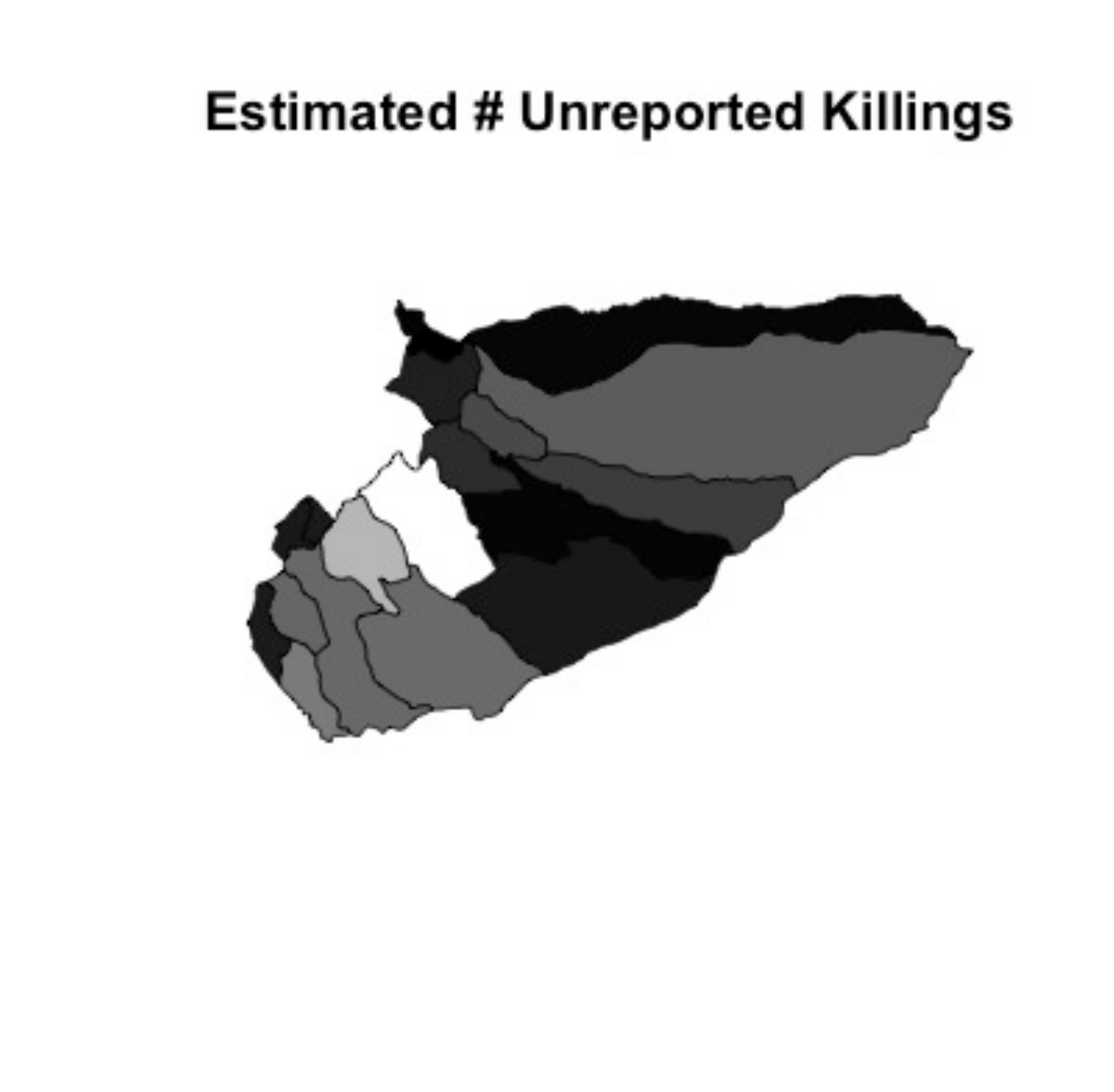}
  \includegraphics[width = 2in]{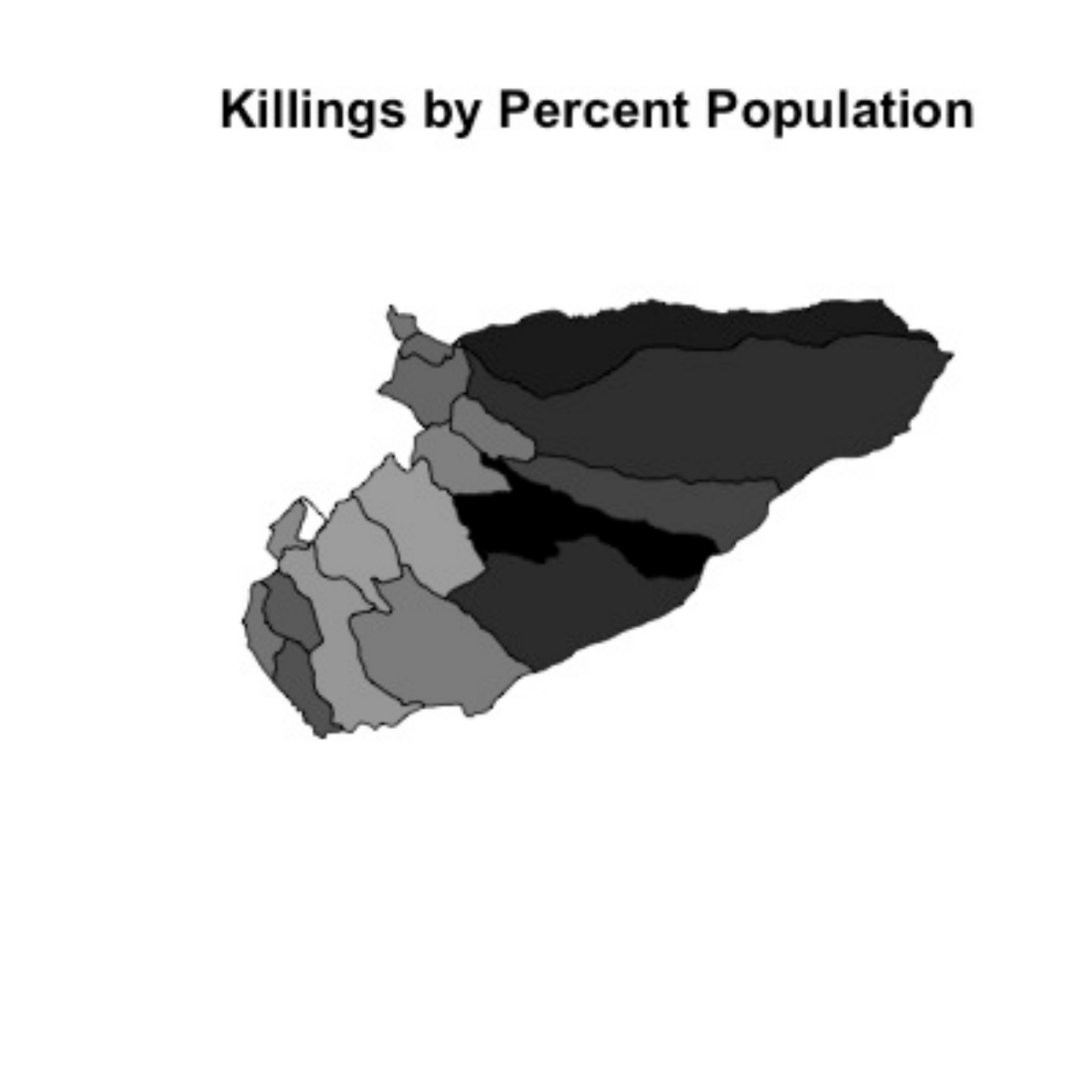}
 \caption{\label{fig:Cas} Posterior mean estimate of the number of unreported killings by region (left) and the mean estimates as a percent of the total population (right).}
 \end{figure}

 We find that from 2001 to 2004, there tended to be an increasing pattern of violence, reaching its peak in this time period in 2004. We also are able to see that the municipalities that experienced the greatest extent of undocumented killings were Aguazul and Yopal, neighboring regions in the middle of the state. These results are shown graphically in Figure \ref{fig:Cas}. Table \ref{tab:CasCompare} provides a comparison of our results to those of \citet{Guberek:2010fk}, labeled ``TCU" due to the title of the article. In this table, we have aggregated our results to match the regional aggregation of \citet{Guberek:2010fk}. The main points of interest in this case is that in almost all region/year strata,  the credible intervals obtained from the SSIP contain the estimates from the previous analysis and vice-versa.  So, despite permitting a much more discretized, detailed view of the pattern of  undocumented killings, this analysis largely confirms the previous results obtained by large-scale spatial aggregation. 

\begin{table}[ht]
\begin{center}
\caption{\label{tab:coltab} Posterior mean and posterior credible intervals of the number of undocumented killings in Casanare, CO by municipality and year.}
\begin{tabular}{rllll}
  \hline
 & 2001 & 2002 & 2003 & 2004 \\ 
  \hline
aguazul & 31 [ 9 , 77 ] & 39 [ 8 , 107 ] & 78 [ 2 , 29 ] & 180 [ 0 , 6 ] \\ 
  chameza & 1 [ 12 , 96 ] & 2 [ 1 , 16 ] & 3 [ 4 , 66 ] & 7 [ 0 , 11 ] \\ 
  hato corozal & 1 [ 24 , 190 ] & 1 [ 1 , 20 ] & 2 [ 0 , 5 ] & 5 [ 1 , 26 ] \\ 
  la salina & 1 [ 57 , 437 ] & 1 [ 3 , 40 ] & 2 [ 0 , 6 ] & 4 [ 1 , 19 ] \\ 
  man\'i & 6 [ 0 , 5 ] & 8 [ 9 , 91 ] & 16 [ 0 , 12 ] & 37 [ 1 , 23 ] \\ 
  monterrey & 6 [ 0 , 6 ] & 7 [ 0 , 7 ] & 15 [ 0 , 26 ] & 34 [ 3 , 46 ] \\ 
  nunchia & 2 [ 0 , 12 ] & 2 [ 0 , 8 ] & 4 [ 0 , 5 ] & 10 [ 8 , 104 ] \\ 
  orocue & 1 [ 1 , 26 ] & 1 [ 0 , 16 ] & 3 [ 0 , 6 ] & 7 [ 0 , 8 ] \\ 
  paz de ariporo & 5 [ 0 , 2 ] & 6 [ 1 , 36 ] & 12 [ 0 , 11 ] & 28 [ 0 , 10 ] \\ 
  por\'e & 3 [ 0 , 3 ] & 4 [ 0 , 4 ] & 8 [ 0 , 24 ] & 19 [ 1 , 20 ] \\ 
  recetor & 1 [ 0 , 5 ] & 2 [ 0 , 5 ] & 3 [ 0 , 3 ] & 7 [ 2 , 45 ] \\ 
  sabanalarga & 1 [ 1 , 10 ] & 2 [ 0 , 10 ] & 3 [ 0 , 4 ] & 7 [ 3 , 31 ] \\ 
  sacama & 1 [ 0 , 3 ] & 1 [ 1 , 23 ] & 2 [ 0 , 7 ] & 4 [ 3 , 38 ] \\ 
  san luis de palenque & 1 [ 0 , 4 ] & 1 [ 1 , 14 ] & 2 [ 0 , 16 ] & 5 [ 7 , 76 ] \\ 
  tamara & 1 [ 0 , 7 ] & 2 [ 1 , 17 ] & 3 [ 0 , 3 ] & 8 [ 18 , 172 ] \\ 
  tauramena & 6 [ 0 , 16 ] & 8 [ 2 , 34 ] & 16 [ 0 , 3 ] & 36 [ 33 , 357 ] \\ 
  trinidad & 2 [ 1 , 19 ] & 3 [ 6 , 77 ] & 6 [ 0 , 6 ] & 14 [ 42 , 448 ] \\ 
  villanueva & 9 [ 1 , 24 ] & 11 [ 0 , 12 ] & 23 [ 1 , 14 ] & 52 [ 85 , 899 ] \\ 
  yopal & 138 [ 3 , 47 ] & 174 [ 1 , 15 ] & 349 [ 0 , 5 ] & 803 [ 200 , 2064 ] \\ 
   \hline
\end{tabular}
\end{center}
\end{table}

\begin{table}[ht]
\begin{center}
\caption{\label{tab:CasCompare}A comparison of the posterior estimates and credible intervals obtained by the Polya-Gamma  regression model with SSIP versus a previous analysis (labeled TCU). }
\begin{tabular}{rllll}
  \hline
 & 2001 & 2002 & 2003 & 2004 \\ 
  \hline
D-SSIP & 174 [ 52 , 401 ] & 218 [ 66 , 503 ] & 437 [ 132 , 1012 ] & 1007 [ 310 , 2317 ] \\ 
  E-SSIP & 7 [ 2 , 17 ] & 9 [ 3 , 21 ] & 18 [ 6 , 41 ] & 42 [ 14 , 92 ] \\ 
  F -SSIP& 30 [ 11 , 66 ] & 38 [ 15 , 83 ] & 75 [ 30 , 165 ] & 173 [ 71 , 375 ] \\ 
  G -SSIP& 14 [ 5 , 30 ] & 17 [ 6 , 37 ] & 35 [ 14 , 73 ] & 79 [ 32 , 166 ] \\ 
  D-TCU & 399 [ 4 , 2198 ] & 365 [ 3 , 2105 ] & 73 [ 0 , 205 ] & 334 [ 4 , 1871 ] \\ 
  E-TCU & NA [ NA , NA ] & 4 [ 0 , 16 ] & 107 [ 0 , 813 ] & 153 [ 0 , 1268 ] \\ 
  F-TCU & NA [ NA , NA ] & NA [ NA , NA ] & 41 [ 0 , 225 ] & 237 [ 2 , 1504 ] \\ 
  G-TCU & 21 [ 0 , 96 ] & 20 [ 0 , 82 ] & 131 [ 1 , 892 ] & 108 [ 3 , 539 ] \\ 
   \hline
\end{tabular}
\end{center}
\end{table}

\section{Discussion and Future Directions} \label{sec:conclusion}
We have introduced a variable selection prior for spatially varying regression that allows for regionally different models, while combining model information via spatially smoothed inclusion probabilities. In any situation in which the liklihood admits a conditionally normal representation, this prior results in a fully conjugate model. We have demonstrated the ability of the SSIP prior improve model selection relative to independent site-specific models in two simulation studies. With improved model selection, we have also shown the potential for improved parameter fitting and prediction in the simulation studies. We applied the SSIP prior to an election dataset and were better able to distinguish sensible regions in which each covariate is explanatory for the percent of the population that voted Democrat in the 2008 presidential election. We show how the SSIP prior can easily be used in a negative binomial regression for capture-recapture. We found that by allowing for regionally varying interaction terms to be included, we were able to reproduce previous results that estimated aggregated regions separately, thus giving us a higher resolution (both spatially and temporally) understanding of the extent of violence in Casanare, Colombia during a period of human rights abuses from 2001-2004. 

In terms of future development of the SSIP, we believe it could  be altered to accommodate the genetic applications such as those described in \citet{Stingo:2011fv}. The SSIP prior in its current form specifies a marginal prior probability of one-half for each covariate. Further work could be done to allow for varying marginal prior inclusion probabilities while maintaining spatial smoothing by placing a prior on the mean of a re-centered latent IAR process.  The zero threshold would remain. For methodological development of the SSIP for CRC, many enhancements could be added to the model suggested above. For example, one could include an offset of the total population in each region to account for different-sized regions or allow separate latent time-series by region. One might also investigate the extent to which spatio-temporal disaggregation is possible under this model.

\bibliography{ssip_20120820_jej_amsart.bbl}

\end{document}